\pdfoutput=1
\documentclass[aps,prb,twocolumn,superscriptaddress]{revtex4-1}

\usepackage{graphicx}
\usepackage{dcolumn}
\usepackage{bm}
\usepackage{amsmath}
\usepackage{amssymb}
\usepackage{latexsym}
\usepackage{epsfig}
\usepackage{amsbsy}
\usepackage{array}
\usepackage{amssymb}
\usepackage{setspace}
\usepackage{bm}
\usepackage{braket}
\usepackage{prettyref}
\usepackage{comment}
\usepackage{mathrsfs}
\usepackage{color,xcolor}
\usepackage{comment}
\usepackage{blkarray}
\usepackage{subfigure}
\usepackage{float}
\usepackage{tikz}

\def\sint{\ifmmode{- \!\!\!\!\!\! \int}
    \else{\hbox{$- \!\!\!\! \int \ $}}\fi}

\newcommand{\reffig}[1]{Fig.~\ref{#1}}

\allowdisplaybreaks
\begin{document}


\title{\textbf{k.p} theory of freestanding narrow band gap semiconductor nanowires}

\author{Ning Luo}
\affiliation{Key Laboratory for the Physics and Chemistry of Nanodevices and Department of Electronics, Peking University, Beijing 100871, China}
\author{Gaohua Liao}
\affiliation{Key Laboratory for the Physics and Chemistry of Nanodevices and Department of Electronics, Peking University, Beijing 100871, China}
\author{H. Q. Xu}\email{hqxu@pku.edu.cn; hongqi.xu@ftf.lth.se}
\affiliation{Key Laboratory for the Physics and Chemistry of Nanodevices and Department of Electronics, Peking University, Beijing 100871, China}
\affiliation{Division of Solid State Physics, Lund University, Box 118, S-221 00 Lund, Sweden}

\date{\today}

\begin{abstract}
We report on a theoretical study of the electronic structures of freestanding nanowires made from narrow band gap semiconductors GaSb, InSb and InAs. The nanowires are described by the eight-band k.p Hamiltonians and the band structures are computed by means of the finite element method in a mixture basis consisting of linear triangular elements inside the nanowires and constrained Hermite triangular elements near the boundaries. The nanowires with two crystallographic orientations, namely the [001] and [111] orientations, and with different cross-sectional shapes are considered. For each orientation, the nanowires of the three narrow band gap semiconductors are found to show qualitatively similar characteristics in the band structures. However, the nanowires oriented along the two different crystallographic directions are found to show different characteristics in the valence bands. In particular, it is found that all the conduction bands show simple, good parabolic dispersions in both the [001]- and [111]-oriented nanowires, while the top valence bands show double-maximum structures in the [001]-oriented nanowires, but single-maximum structures in the [111]-oriented nanowires. The wave functions and spinor distributions of the band states in these nanowires are also calculated. It is found that significant mixtures of electron and hole states  appear in the bands of these narrow band gap semiconductor nanowires. The wave functions exhibit very different distribution patterns in the nanowires oriented along the [001] direction and the nanowires oriented along the [111] direction. It is also shown that single-band effective mass theory could not reproduce all the band state wave functions presented in this work.
\end{abstract}
\maketitle

\section{Introduction}  

Recently semiconductor nanowires have attracted great attention due to their unique physical properties and potential applications in nanoelectronics, optoelectronics, and quantum electronics.\cite{Yang2014,Magnusson2014,Riel2011,Chung-1,Kimberly-1,Caroff-1,Boxberg2010,Borg-1,Ganjipour2011,Thelander-1,Xu-9,Nilsson2010,Nilsson2011,Mourik2012,Xu-8,Churchill2013,Deng2014} With advances in the materials technology, high-quality semiconductor nanowire have been obtained through, for instance, molecular-beam epitaxy,\cite{de2013review,liu2005artificial,hertenberger2010growth} metal-organic vapor phase epitaxy,\cite{caroff2009insb,jeppsson2008characterization,haas2013nanoimprint} and chemical vapor deposition.\cite{joyce2007growth,guvenc2011data,guo2006structural,MingLi2015,KanLi2016-1,KanLi2016-2} Due to their well organized crystal structures, relatively high carrier mobilities, small cross sections, and strong quantum confinement effects,  III-V semiconductor nanowires have been employed to construct field-effect transistors,\cite{cui2003high,paul2010chemical,jing2009insb,jia2011field} infrared photodetectors,\cite{chen2009infrared,liu2013high} light emission diodes,\cite{nguyen2011p,qian2005core} thermal electrical devices,\cite{boukai2008silicon,hochbaum2008enhanced} laser devices,\cite{duan2003single,huang2001room} solar cells,\cite{tian2007coaxial,colombo2009gallium,krogstrup2013single,Wallentin2013} and quantum devices.\cite{Xu-9,Nilsson2010,Nilsson2011,Mourik2012,Xu-8,Churchill2013,Deng2014,Nilsson2012,Abay2012,Abay2013,Abay2014}

Several theoretical methods, including density-functional theory,\cite{ning2013first,cahangirov2009first,srivastava2013first,dos2010diameter} tight-binding methods,\cite{persson2008electronic,persson2004electronic,persson2006electronic,persson2006electronic2,niquet2007effects,niquet2006electronic,luisier2006atomistic,2015Liao,2016Liao,2016Liao2} and k.p theory,\cite{2006Lassen111InPInAs,kohn1955theory,luttinger1956quantum,citrin1989valence,xia1991effective,2012PeetersInAs/GaSb,2012Peeters6v8,lassen2004exact,2006Lassen111InPInAs,Peeters2008optical,kishore2010electronic} have been used to study the properties of semiconductor nanowires. Among all the methods, k.p theory is considered as a computationally less-demanding one and is wildly adopted in the calculations for the nanowire band structures near the band extrema. Both narrow and large band gap nanowires grown along the [001] crystallographic direction have been studied based on $6\times 6$ or $8\times 8$ k.p Hamiltonians.\cite{2012Peeters6v8,2012PeetersInAs/GaSb,kishore2010electronic}
However, few studies have been made for the properties of more frequently experimentally grown [111]-oriented nanowires based on k.p theory. Redi\'nski and Peeters have calculated the band structure of GaAs nanowires oriented in the [111] direction based on a six-band k.p Hamiltonian.\cite{Peeters2008optical}
Lassen {\em et al.} have made a study of the band structure of [111]-oriented GaAs nanowires based on a $4\times 4$ Burt-Foreman Hamiltonian\cite{lassen2004exact} and a study of the band structures of [111]-oriented InP and InAs nanowires based on $8\times 8$ Hamiltonians\cite{2006Lassen111InPInAs}.
However, the studies by Lassen {\em et al.} have only be made for [111]-oriented InP and InAs nanowires with  small cross-sectional sizes of 2.5 nm to 15 nm and are lack of the analysis of the conduction band  properties and the spinor distributions in the band states, which can play an essential role in the understanding of the electronic properties of nanowires, especially, for those made from narrow band gap semiconductors\cite{2012PeetersInAs/GaSb}. Thus, the capability of k.p theory have yet to be fully explored and extended to [111]-oriented nanowires with large cross-sectional sizes, which are seldom discussed theoretically, but are more commonly grown and employed in experiments.

The purpose of this paper is to present a systematic study of the electronic structures of narrow band gap semiconductor GaSb, InSb and InAs nanowires oriented along the [001] and [111] crystallographic directions based on  $8\times 8$ k.p theory. A form of an eight-band Luttinger-Kohn Hamiltonian with a principal axis along the [111] direction has been derived. Such a Hamiltonian has not yet been present before, but it is required to study the electronic structure of nanowires grown along the [111] crystallographic direction. The paper is organized as follows. In Section II, the eight-band k.p theory applicable to nanowires oriented in the [001] and [111] directions and the choices for materials parameters are presented and discussed.  The methods of numerical computations have also been briefly described in this section, while the details are presented and discussed in the Appendix. In Section III, the results of calculations for [001]-oriented GaSb, InSb and InAs nanowires based on the standard eight-band Luttinger-Kohn Hamiltonian are presented and discussed. In Section IV, the results of calculations for [111]-oriented GaSb, InSb and InAs nanowires based on the derived form of the eight-band Luttinger-Kohn Hamiltonian as shown in Section II are presented and discussed, and are compared with the  corresponding results obtained for the [001]-oriented nanowires. Finally, the paper is summarized in Section V.

\section{Theory}  

In this section, we present the k.p theory employed in this work. For the study of narrow band gap semiconductor nanowires, it becomes very important to adopt a theory that includes the interaction between the conduction bands and the valence bands. In this work, an eight-band Luttinger-Kohn Hamiltonian\cite{bahder1990eight} will be adopted. For narrow band gap semiconductor GaSb, InSb and InAs nanowires oriented in the [001] crystallographic direction, the standard form of the Hamiltonian can be employed directly after replacing the $k_x$ and $k_y$ vectors with the corresponding momentum operators. However, for the treatment of experimentally more commonly grown, [111]-oriented GaSb, InSb and InAs nanowires, one needs to transform the standard form of the eight-band Luttinger-Kohn Hamiltonian to the form with the [111] direction as a principal axis. In the following, a derivation of an eight-band Luttinger-Kohn Hamiltonian with an arbitrary crystallographic direction will be presented and will then be applied to obtain the form which can be readily used in the calculations for an nanowire oriented along the [111] direction. The numerical implementation and the choices of the parameters in the treatment of GaSb, InSb and InAs nanowires will also be described and discussed.

\subsection{Derivation of an eight-band Hamiltonian with its principal axis in an arbitrary direction}

Our starting point in deriving a form of the eight-band Luttinger-Kohn Hamiltonian with an arbitrary crystallographic direction as a principal axis is a standard $4\times 4$ Hamiltonian matrix,\cite{kane1957band} $H_{4\times 4}$, widely used in the study of direct band gap semiconductors. The matrix is in the form of
\begin{widetext}
\small
\begin{eqnarray}
\begin{blockarray}{cccc}
\ket{S}&\ket{X}&\ket{Y}&\ket{Z}\\
\begin{block}{[cccc]}
E_c+(A'+\frac{\hbar^2}{2m_0})k^2&Bk_yk_z+iP_0k_x&Bk_xk_z+iP_0k_y&Bk_xk_y+iP_0k_z\\
Bk_yk_z-iP_0k_x&E'_v+M(k_y^2+k_z^2)+L'k_x^2+\frac{\hbar^2k^2}{2m_0}&N'k_xk_y&N'k_xk_z\\
Bk_xk_z-iP_0k_y&N'k_xk_y&E'_v+M(k_x^2+k_z^2)+L'k_y^2+\frac{\hbar^2k^2}{2m_0}&N'k_yk_z\\
Bk_xk_y-iP_0k_z&N'k_xk_z&N'k_yk_z&E'_v+M(k_x^2+k_y^2)+L'k_z^2+\frac{\hbar^2k^2}{2m_0}\\
\end{block}
\end{blockarray}\; ,
\label{eq:hamiltonian4}
\end{eqnarray}
\normalsize
\end{widetext}
where $E_c$ and $E'_v$ are the bulk conduction and valence band wdge energies at the $\Gamma$-point ($\textbf{\textit{k}}=0$), $P_0$ describes interaction between the conduction and the valence bands, $L'$, $M$, and $N'$ are parameters related closely to valence band structures, $A'$ and $B$ come from the second order interactions involving the states outside the considered conduction and valence bands. The basis is chosen to be the ordinary set of $\{\ket{S},\ket{X},\ket{Y},\ket{Z}\}$ as denoted on the top of the matrix in Eq.~\eqref{eq:hamiltonian4}.

The eight-band k.p Luttinger-Kohn Hamiltonian with the [001] direction as a principal axis could be derived based on the above $4\times 4$ Hamiltonian matrix and can be expressed in terms of a spin-product basis of $\{\alpha\}$=$\{\ket{S\downarrow}$,$\ket{X\downarrow}$,$\ket{Y\downarrow}$,$\ket{Z\downarrow}$, $\ket{S\uparrow}$,$\ket{X\uparrow}$,$\ket{Y\uparrow}$,$\ket{Z\uparrow}\}$ as
\begin{eqnarray}
H=H^0+H^{so},
\label{eq:full}
\end{eqnarray}
where $H^0$ is the $8\times 8$ matrix constructed from Eq.~(\ref{eq:hamiltonian4i}),
\begin{eqnarray}
H^0=\begin{bmatrix}H_{4\times 4}&\\&H_{4\times 4}\end{bmatrix},
\label{eq:h0def}
\end{eqnarray}
and $H^{so}$ is the spin-orbit interaction matrix,
\begin{eqnarray}
H^{so}=\frac{\Delta}{3}\begin{bmatrix}
1&0&0&0&0&0&0&0\\
0&0&i&0&0&0&0&-1\\
0&-i&0&0&0&0&0&-i\\
0&0&0&0&0&1&i&0\\
0&0&0&0&1&0&0&0\\
0&0&0&1&0&0&-i&0\\
0&0&0&-i&0&i&0&0\\
0&-1&i&0&0&0&0&0\\
\end{bmatrix}.
\label{eq:hsodef}
\end{eqnarray}
The complex matrix $H^{so}$  could be diagonalized in the basis of \{$\ket{JM}$\}, where $\ket{JM}$ represents a state with the total angular momentum quantum number $J=1/2$ or $3/2$ and the component quantum number $M$, leaving a value of $\Delta/3$ added to the heavy-hole (HH) and the light-hole (LH) states and a value of $-2\Delta/3$ added to the spin split-off (SO) state.\cite{bahder1990eight} In the Luttinger-Kohn formulation, the basis set of \{$\ket{JM}$\} is often chosen to present the eight-band k.p Hamiltonian. In the treatment of nanowires oriented in the [001] direction, the eight-band Hamiltonian as in Eq.~\eqref{eq:full} or its corresponding form represented in the basis set of \{$\ket{JM}$\} needs to be solved after replacing $k_x$ and $k_y$ with the corresponding momentum operators.


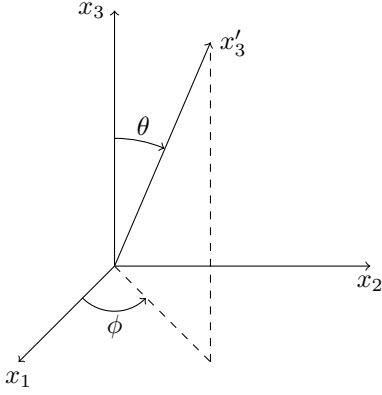
\begin{figure}[t]
\centering
\begin{tikzpicture}[scale=0.85]
\draw[->] (0,0) -- (4,0) node[below] {$x_2$};
\draw[->] (0,0) -- (0,4) node[left] {$x_3$};
\draw[->] (0,0) -- (-1.5,-1.5) node[below] {$x_1$};
\draw[->] (0,0) -- (1.5,3.5) node[right] {$x'_3$};
\draw[dashed] (0,0) -- (1.5,-1.5);
\draw[dashed] (1.5,-1.5) -- (1.5,3.5);
\draw[->] (0,2) arc (90:67:2);
\draw[->] (-0.5,-0.5) arc(225:315:0.7);
\node at (0.45,2.2) {$\theta$};
\node at (0,-0.95) {$\phi$};
\end{tikzpicture}
\caption{Illustration for coordinate systems where the new coordinate system is set with its principal axis $x^{\prime}_3$ being rotated by azimuth angle ($\theta$,$\phi$) from the original coordinate system.}
\label{fig:azimuth}
\end{figure}

In the treatment of narrow band gap semiconductor nanowires oriented in a direction other than [001] direction, it is however convenient to adopt an eight-band k.p Hamiltonian described with a principal axis along the direction. Such a Hamiltonian could be derived from a standard Luttinger-Kohn eight-band Hamiltonian using a rotation operator,\cite{2006Lassen111InPInAs}
\begin{eqnarray}
R(\phi,\theta)=\begin{pmatrix}
\cos(\phi)\cos(\theta)&\sin(\phi)\cos(\theta)&-\sin(\theta)\\
-\sin(\phi)&\cos(\phi)&0\\
\cos(\phi)\sin(\theta)&\sin(\phi)\sin(\theta)&\cos(\theta)
\end{pmatrix},
\label{eq:5Rotate}
\end{eqnarray}
where $\theta$ and $\phi$ define the azimuth angle as illustrated in \reffig{fig:azimuth}. Under the rotation operation, the original axis systems are transformed to the desired ones according to
\begin{eqnarray}
x'_i=R_{ij}x_j,\quad k'_i=R_{ij}k_j .
\label{eq:6Rotate}
\end{eqnarray}
The orbital states in the basis set of $\{\beta\}$=$\{\ket{S'\downarrow}$, $\ket{X'\downarrow}$, $\ket{Y'\downarrow}$, $\ket{Z'\downarrow}$, $\ket{S'\uparrow}$, $\ket{X'\uparrow}$, $\ket{Y'\uparrow}$, $\ket{Z'\uparrow}\}$ are connected with the orbital states in the original basis set by
\begin{eqnarray}
\{\beta\}=U\{\alpha\},\quad U=
\begin{bmatrix}
1&&&\\
&R&&\\
&&1&\\
&&&R
\end{bmatrix},
\label{eq:7Rotate}
\end{eqnarray}
and the spin states are transformed according to
\begin{eqnarray}
\begin{bmatrix}\downarrow'\\\uparrow'\end{bmatrix}=\begin{bmatrix}e^{i\frac{\phi}{2}}\cos(\frac{\theta}{2})&-e^{-i\frac{\phi}{2}}\sin(\frac{\theta}{2})\\
e^{i\frac{\phi}{2}}\sin(\frac{\theta}{2})&e^{-i\frac{\phi}{2}}\cos(\frac{\theta}{2})\end{bmatrix}\begin{bmatrix}\downarrow\\\uparrow\end{bmatrix}.
\label{eq:8Rotate}
\end{eqnarray}
Thus, the basis set defined with the principal axes being set along the axes of the rotated system,
$\{\gamma\}$=$\{\ket{S'\downarrow'}$, $\ket{X'\downarrow'}$, $\ket{Y'\downarrow'}$, $\ket{Z'\downarrow'}$, $\ket{S'\uparrow'}$, $\ket{X'\uparrow'}$, $\ket{Y'\uparrow'}$, $\ket{Z'\uparrow'}\}$, are connected to the original basis set $\{\alpha\}$ as
\begin{eqnarray}
\{\gamma\}=A U\{\alpha\}= W\{\alpha\},
\end{eqnarray}
where
\begin{eqnarray}
A=\begin{bmatrix}e^{i\frac{\phi}{2}}\cos(\frac{\theta}{2})I_4&-e^{-i\frac{\phi}{2}}\sin(\frac{\theta}{2})I_4\\
e^{i\frac{\phi}{2}}\sin(\frac{\theta}{2})I_4&e^{-i\frac{\phi}{2}}\cos(\frac{\theta}{2})I_4\end{bmatrix},
\end{eqnarray}
with $I_4$ is a $4\times 4$ identity matrix. The Hamiltonian matrix $H_{\gamma}$ expressed in terms of the basis set \{$\gamma$\} is related to the k.p Hamiltonian matrix $H_{\alpha}$, which is expressed in terms of the original basis set \{$\alpha$\}, via
\begin{eqnarray}
H_{\gamma}=W^*H_{\alpha}W^T.
\label{eq:basis-rotation}
\end{eqnarray}
In the Hamiltonian matrix given in Eq.~\eqref{eq:h0def}, each element  is a polynomial of \textbf{\textit{k}} vector components up to the second power. Here \textbf{\textit{k}} is defined in the original coordinate system. In order to conveniently treat nanowires oriented along an arbitrary direction defined by azimuth angle ($\theta, \phi$), we would like to define the  \textbf{\textit{k}} vector in the Hamiltonian in the rotated system. In terms of this newly defined $\textbf{\textit{k}}^{\prime}$ vector, the original Hamiltonian $H^0_\alpha$ is transformed according to
\begin{eqnarray}
H^0_\alpha(\textbf{k}')=H^0_\alpha(R^{-1}\textbf{k}).
\label{eq:k-rotation}
\end{eqnarray}
By combining Eqs.~\eqref{eq:basis-rotation} and \eqref{eq:k-rotation}, the Hamiltonian $H^0$ defined in terms of the basis set \{$\gamma$\} with a principal axis along the direction set by azimuth angle ($\theta,\phi$) can be found from
\begin{eqnarray}
H^0_\gamma(\textbf{k}')=W^* H^0_\alpha(R^{-1}\textbf{k}) W^T.
\label{eq:9Rotate}
\end{eqnarray}
The result shown in Eq.~\eqref{eq:9Rotate} is a concise form without including the spin-orbit interaction term $H^{so}$ given in Eq.~\eqref{eq:hsodef}.  This term is of a diagonal form in the basis of \{$\delta$\}=\{$\ket{J^{\prime}M^{\prime}}$\}.  Here we use $\ket{J^{\prime}M^{\prime}}$ to denote that the state angular momenta are defined with the respect to the principal axes in the rotated coordinate systems. It can be shown that
the basis set \{$\delta$\} is connected to the basis set \{$\gamma$\} via
\begin{eqnarray}
\{\delta\}=Q\{\gamma\},
\end{eqnarray}
where the transformation matrix $Q$ is given by
\begin{eqnarray}
Q=\begin{bmatrix}
1&0&0&0&0&0&0&0\\
0&0&0&0&1&0&0&0\\
0&-\frac{\sqrt{6}}{6}i&\frac{\sqrt{6}}{6}&0&0&0&0&\frac{\sqrt{6}}{3}i\\
0&0&0&0&0&\frac{\sqrt{2}}{2}i&-\frac{\sqrt{2}}{2}&0\\
0&-\frac{\sqrt{2}}{2}i&-\frac{\sqrt{2}}{2}&0&0&0&0&0\\
0&0&0&\frac{\sqrt{6}}{3}i&0&\frac{\sqrt{6}}{6}i&\frac{\sqrt{6}}{6}&0\\
0&0&0&\frac{\sqrt{3}}{3}i&0&-\frac{\sqrt{3}}{3}i&-\frac{\sqrt{3}}{3}&0\\
0&-\frac{\sqrt{3}}{3}i&\frac{\sqrt{3}}{3}&0&0&0&0&-\frac{\sqrt{3}}{3}i\\
\end{bmatrix}.
\end{eqnarray}
The eight-band k.p Luttinger-Kohn Hamiltonian with the principal axes defined in the rotated coordinate system can then be expressed in terms of \{$\delta$\} as
\begin{eqnarray}
H^0_\delta(\textbf{k}')=&&P^*H^0_\alpha(R^{-1}\textbf{k})P^T\nonumber\\
&&+diag\{0,0,\frac{\Delta}{3},\frac{\Delta}{3},\frac{\Delta}{3},\frac{\Delta}{3},-\frac{2\Delta}{3},-\frac{2\Delta}{3}\},
\end{eqnarray}
where
\begin{eqnarray}
P=QAU. \label{eq:general}
\end{eqnarray}
Equation \eqref{eq:general} presents a general expression for transformation of the standard eight-band k.p Luttinger-Kohn Hamiltonian to an eight-band k.p Luttinger-Kohn Hamiltonian defined in the basis set of \{$\delta$\} with the principal axes set along the axes of a rotated coordinate system of an arbitrary rotation angle  ($\theta, \phi$).

By setting the values of $(\theta,\phi)$ to $[\arccos (1/\sqrt{3}), \pi/4]$ and  the $x'_3$ axis to  [111] direction, we have the eight-band k.p Luttinger-Kohn Hamiltonian matrix as
\begin{widetext}
\small
\begin{eqnarray}
\begin{bmatrix}
A&0&\bar{T}+\bar{V}&0&-\sqrt{3}(T-V)&\sqrt{2}(W-{U})&W-U&\sqrt{2}(\bar{T}+\bar{V})\\
0&A&\sqrt{2}(W-{U})&-\sqrt{3}(\bar{T}+\bar{V})&0&T-V&-\sqrt{2}(T-V)&\bar{W}+U\\
T+V&\sqrt{2}(\bar{W}-U)&-{P}+Q&-\bar{S}&R&0&\sqrt{\frac{3}{2}}S&-\sqrt{2}Q\\
0&-\sqrt{3}(T+V)&-S&-P-Q&0&R&-\sqrt{2}R&\frac{1}{\sqrt{2}}S\\
-\sqrt{3}(\bar{T}-\bar{V})&0&\bar{R}&0&-P-Q&\bar{S}&\frac{1}{\sqrt{2}}\bar{S}&\sqrt{2}\bar{R}\\
\sqrt{2}(\bar{W}-U)&\bar{T}-\bar{V}&0&\bar{R}&S&-P+Q&\sqrt{2}Q&\sqrt{\frac{3}{2}}\bar{S}\\
\bar{W}-U&-\sqrt{2}(\bar{T}-\bar{V})&\frac{\sqrt{6}}{2}\bar{S}&-\sqrt{2}\bar{R}&\frac{\sqrt{2}}{2}S&\sqrt{2}Q&Z&0\\
\sqrt{2}(T+V)&W+U&-\sqrt{2}Q&\frac{\sqrt{2}}{2}\bar{S}&\sqrt{2}R&\frac{\sqrt{6}}{2}S&0&Z\\
\end{bmatrix},
\label{eq:hamiltonian111}
\end{eqnarray}
\end{widetext}
\normalsize
where the parameters are defined as in Ref.~\onlinecite{bahder1990eight} except for some specific expressions,
\begin{eqnarray*}
A&=&E_c+(A'+\frac{\hbar^2}{2m_0})(k^2_x+k^2_y+k^2_z),\\
U&=&\frac{1}{\sqrt{3}}P_0k_z ,\\
V&=&\frac{1}{\sqrt{6}}P_0(k_x-ik_y),\\
{W}&=&-\frac{i}{6}B(k_x^2+k_y^2-2k_z^2),\\
{T}&=&-\frac{i}{6}B(k_x+ik_y)^2-\frac{\sqrt{2}i}{6}Bk_z(k_x-ik_y),\\
P&=&-E_v+\tilde{\gamma_1}\frac{\hbar^2}{2m_0}(k^2_x+k^2_y+k^2_z),\\
{Q}&=&\tilde{\gamma_3}\frac{\hbar^2}{2m_0}(k^2_x+k^2_y-2k^2_z),\\
{R}&=&-\frac{\sqrt{3}}{6}\frac{\hbar^2}{m_0}(\tilde{\gamma_2}+2\tilde{\gamma_3})(k_x-ik_y)^2,\\
&&+\frac{\sqrt{6}}{3}\frac{\hbar^2}{m_0}(\tilde{\gamma_2}-\tilde{\gamma_3})k_z(k_x+ik_y),\\
{S}&=&-\frac{\sqrt{6}}{6}\frac{\hbar^2}{m_0}(\tilde{\gamma_2}-\tilde{\gamma_3})(k_x+ik_y)^2,\\
&&+\frac{\sqrt{3}}{3}\frac{\hbar^2}{m_0}(2\tilde{\gamma_2}+\tilde{\gamma_3})k_z(k_x-ik_y),\\
Z&=&E_v-\Delta-\tilde{\gamma_1}\frac{\hbar^2}{2m_0}(k_x^2+k_y^2+k_z^2),
\end{eqnarray*}
with the top bulk valence band energy at the $\Gamma$-point, $E_v$, given by
\begin{equation}
E_v=E_{v'}+\frac{\Delta}{3}.
\end{equation}
In the above expressions, parameters $\tilde{\gamma_1}$, $\tilde{\gamma_2}$ and $\tilde{\gamma_3}$ are the modified Luttinger parameters\cite{1966PB} which are related to the Dresselhaus parameters\cite{dresselhaus1955cyclotron} $L'$,$M$ and $N'$ by
\begin{eqnarray}
\tilde{\gamma_1}&=&-\frac{2}{3}\frac{m_0}{\hbar^2}(L'+2M)-1 ,\nonumber\\
\tilde{\gamma_2}&=&-\frac{1}{3}\frac{m_0}{\hbar^2}(L'-M) ,\\
\tilde{\gamma_3}&=&-\frac{1}{3}\frac{m_0}{\hbar^2}N' . \nonumber
\end{eqnarray}
In the present work, these modified Luttinger parameters $\tilde{\gamma_1}$, $\tilde{\gamma_2}$, and $\tilde{\gamma_3}$ are assumed to take the forms of Pidgeon and Brown\cite{1966PB}, instead of the forms of Bahder\cite{ bahder1990eight} and Pryor\cite{pryor1998eight}, as suggest by Kishore {\em et al.}\cite{2012Peeters6v8},
\begin{eqnarray}
\tilde{\gamma_1}=\gamma_1-\frac{E_p}{3E_g}, \nonumber\\
\tilde{\gamma_2}=\gamma_2-\frac{E_p}{6E_g}, \\
\tilde{\gamma_3}=\gamma_3-\frac{E_p}{6E_g}, \nonumber
\end{eqnarray}
where $E_g$ is the bulk band gap, $E_p$ the Kane energy parameter, and $\gamma_1$, $\gamma_2$ and $\gamma_3$  the Luttinger parameters.

\subsection{Choice of parameters}

In the calculations for this work, most of the band and Luttinger parameters are taken from Vurgaftman \textit{et al.}\cite{vurgaftman2001band}, except for parameters $E_p$ for GaSb and InSb. As suggested by Foreman\cite{foreman1997elimination}, we have taken a redefined form of $E_p$, to eliminate spurious solutions from the numerical calculations for GaSb and InSb nanowires, as
\begin{eqnarray}
E_p=\frac{3m_e/m_c}{2/E_g+1/(E_g+\Delta)},
\label{eq:E_p}
\end{eqnarray}
where $m_e$ is the free electron mass and $m_c$ the conduction-band effective electron mass. We adopt this equation to eliminate the spurious solution. The parameters used to calculated $E_p$ are taken from Ref.~\onlinecite{vurgaftman2001band}, except for the $m_c$ for bulk InSb. In the calculations for InSb nanowires, we still face problems with spurious solutions even after we have modified parameter $E_p$ using the value of $m_c$ recommended by Vurgaftman \textit{et al.} in Ref.~\onlinecite{vurgaftman2001band}. Thus, in the treatment of InSb nanowires, we have used  the value of $m_c$ as suggested  for bulk InSb in Ref.~\onlinecite{johnson1970infrared}, which is still in the reasonable range as suggested by Vurgaftman \textit{et al.} Table \ref{tab:1} summarizes all the required materials parameters used in the calculations for this work, except for parameters $A'$ and $B$. These two parameters have been set to zero in the calculations. In this work, we have also taken the bulk valence band edges as the energy reference by setting $E_v=0$ in all the calculations.

\begin{table}[tb]
\caption{Band and Luttinger parameters for bulk materials InAs, GaSb and InSb used in this work. All the parameters are taken from Ref.~\onlinecite{vurgaftman2001band}, except for $m_c$ for bulk InSb. This parameter is taken from Ref.~\onlinecite{johnson1970infrared}.}
\label{tab:1}
\begin{tabular}{lcccc}
\hline
\hline
&  & InAs & GaSb & InSb \\
\hline
& $E_g$ (eV) & \; 0.417 \; & \; 0.812 \; & \; 0.235\; \\
& $\Delta$ (eV) & 0.39 & 0.76 & 0.81\\
& $E_p$ (eV) & 21.5 & 24.76 & 23.2\\
& $\gamma_1$ & 20.0 & 13.4 & 34.8\\
& $\gamma_2$ & 8.5 & 4.7 & 15.5\\
& $\gamma_3$ & 9.2 & 6.0 & 16.5\\
&$a_0 (nm)$&0.61&0.61&0.65\\
&$m_c/m_e$&     &0.039&0.0139\\
\hline
\end{tabular}
\end{table}

\subsection{Numerical implementation}

\begin{figure}[tb]
\includegraphics[width=7cm]{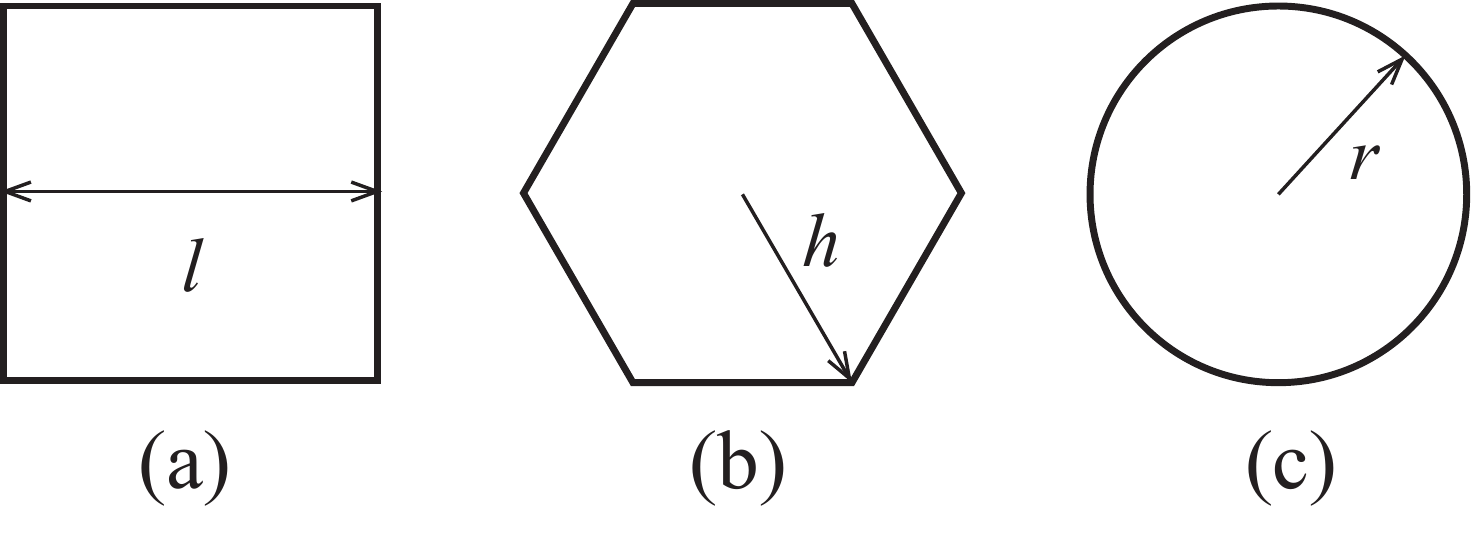}
\caption{Cross sections of nanowires considered in this work. The sizes of square, hexagonal and circular cross sections are measured by side length \textit{l}, side length \textit{h} and radius \textit{r}, respectively.\label{fig:Cross_section_shape}}
\end{figure}

Finite element method (FEM) is a numerical technique to find the solutions of partial differential equations. Here we will briefly introduce the numerical procedure in solving for the eigensolutions of our nanowire Hamiltonians using FEM. For more details about the procedure, we would like to refer to the Appendix. In the implement of FEM in this work, the nanowire eigenvalue equation,
\begin{equation}
HF=EF,
\end{equation}
where $H$ is the Hamiltonian and $F$ is the envelop function,
is discretized in a mixture basis consisting linear triangular elements inside the cross section and constrained Hermite triangular elements near the boundaries. Three nanowire cross-sectional shapes, namely, square, hexagon, and circle, have been considered in this work and the cross-sectional  sizes of the nanowires are measured 
by side length \textit{l}, side length \textit{h} and radius \textit{r}, respectively, see the schematics shown in \reffig{fig:Cross_section_shape}. The inclusion of high-order constrained Hermite basis has significantly suppressed the Gibbs oscillations in the obtained envelop functions\cite{2005SenguptaGibbs} as shown in Section~\ref{part:001} and Section~\ref{part:111}. 

The spinor distribution analysis is useful in understanding the characteristics of band states in a nanowire. In an eight-band k.p theory, the spinor distributions in a band state of a nanowire, namely the contributions from bulk electron (EL), heavy-hole (HH), light-hole (LH) and spin split-off (SO) states are of interest, which we will denote by real numbers of $C_{el}$, $C_{hh}$, $C_{lh}$ and $C_{so}$ with
\begin{equation}
C_{el}+C_{hh}+C_{lh}+C_{so}=1.
\end{equation}
The spinor distributions can readily be evaluated out from the obtained envelope functions since they are presented in terms of \{$|JM\rangle$\} or \{$|J^{\prime}M^{\prime}\rangle$\} in this work as we have discussed above.

\section{Band structures and band states of [001]-oriented G\lowercase{a}S\lowercase{b}, I\lowercase{n}A\lowercase{s} and I\lowercase{n}S\lowercase{b} nanowires\label{part:001}}

In the following, we first present our results of calculations for [001]-oriented GaSb, InAs, and InSb nanowires. 
Since the Rashbar or Dresselhaus term is not included in our k.p Hamiltonians, the energy bands of these nanowires are all doubly spin-degenerated. Thus, for each band state, we will only present the spatial distribution of the wave function  of one spin state since the spatial distribution for the other one is identical.

\subsection{GaSb and InSb nanowires with a square cross section}


\begin{figure}[tb]
\includegraphics[width=8.5cm]{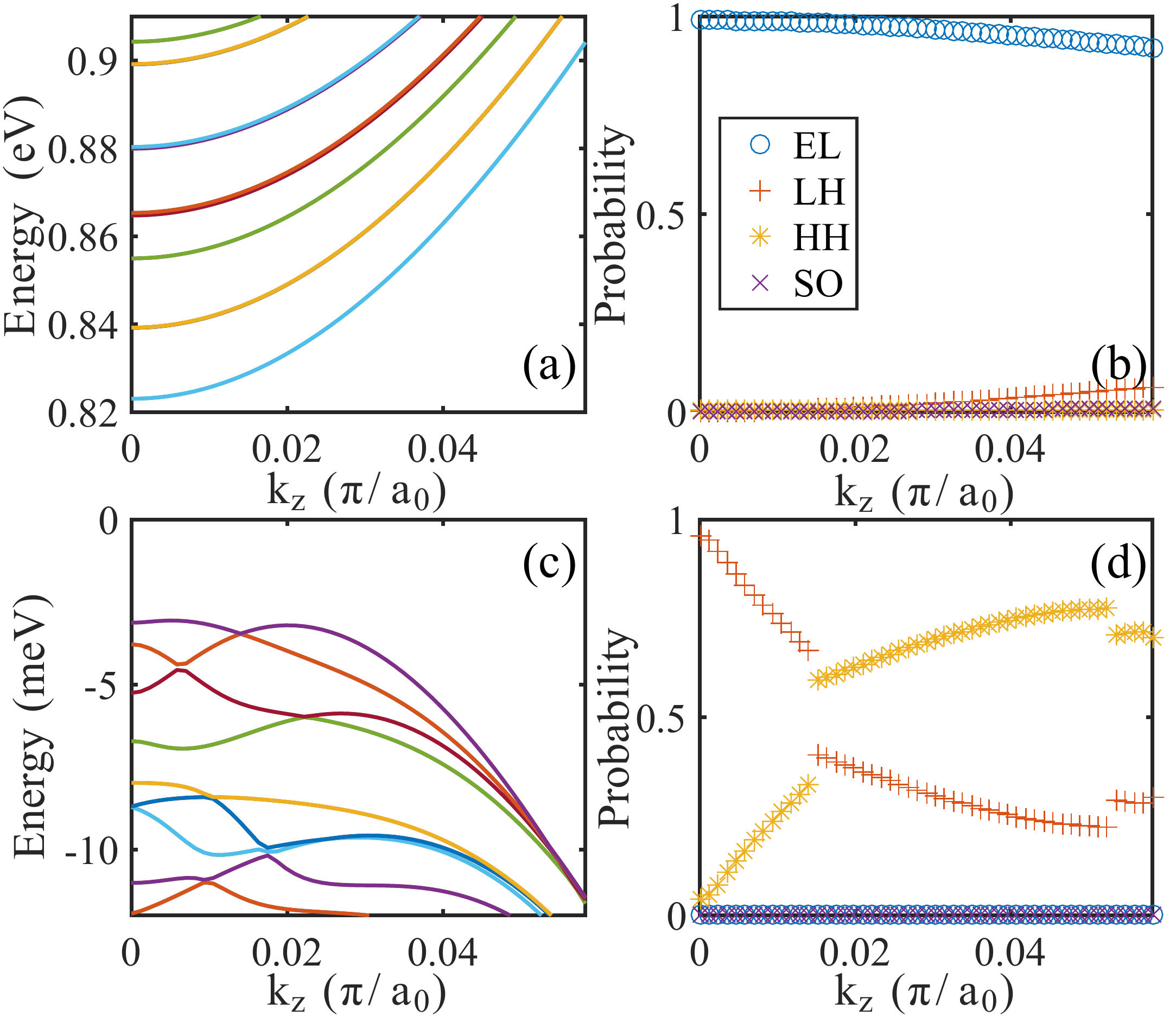}
\caption{(a) and (c) Band structure of a GaSb nanowire oriented along the [001] crystallographic direction with a square cross section of size $l=40$ nm. (b) and (d) Spinor distributions of the lowest conduction band state and the highest valence band state at the $\Gamma$ point of the GaSb nanowire. 
\label{fig:f001GaSb_Band_Spinor}}
\end{figure}

\begin{figure}[tb]
\includegraphics[width=8.5cm]{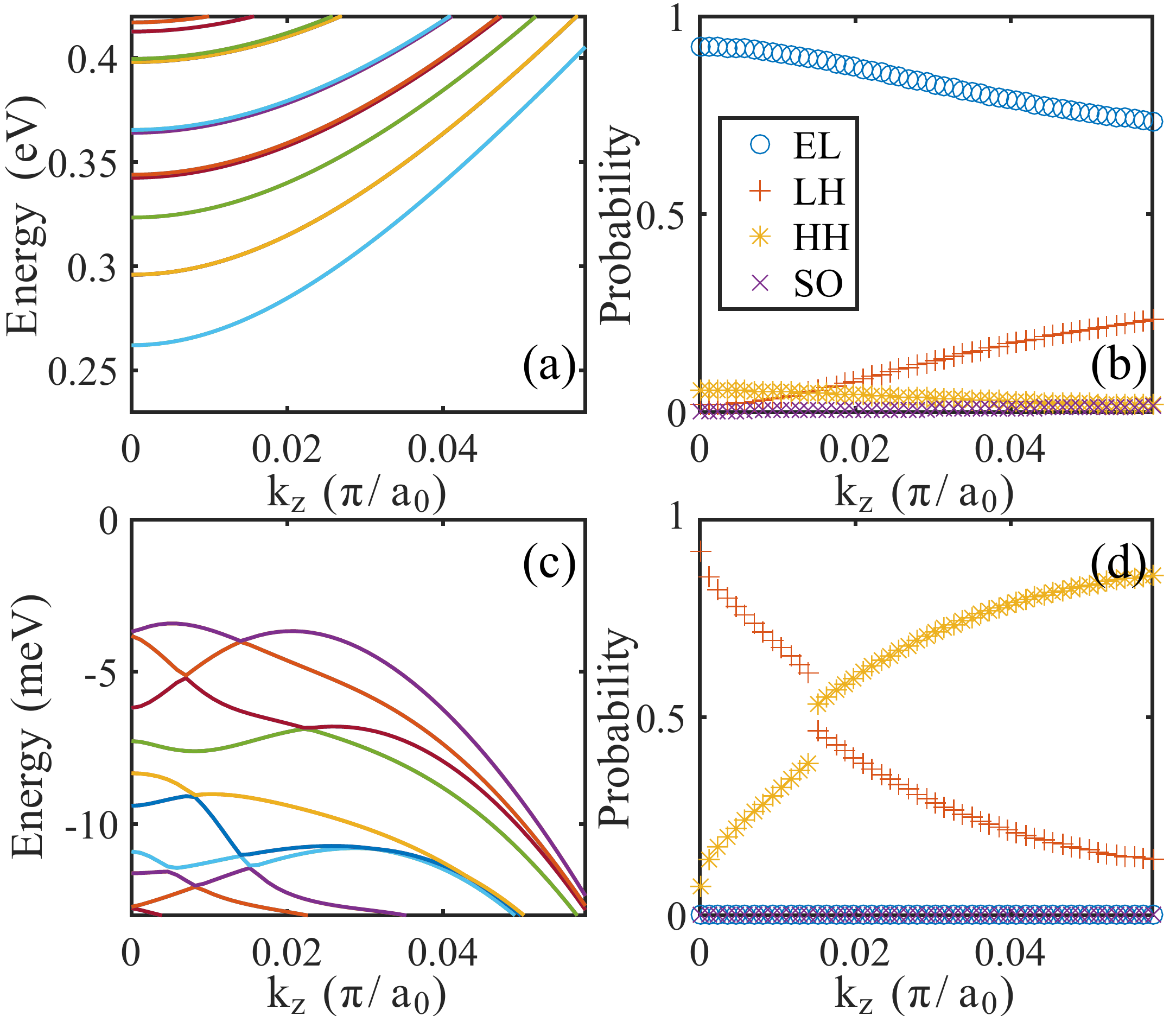}
\caption{(a) and (c) Band structure of an InSb nanowire oriented along the [001] crystallographic direction with a square cross section of size $l=40$ nm. (b) and (d) Spinor distributions of the lowest conduction band state and the highest valence band state at the $\Gamma$ point of the InSb nanowire. 
\label{fig:f001InSb_Band_Spinor}}
\end{figure}

Let us first consider the band properties of [001]-oriented GaSb and InSb nanowires with a square cross section. Among the three narrow band gap III-V semiconductors, GaSb has the largest bulk band gap and the InSb the smallest bulk band gap (cf.~Table~\ref{tab:1}). Thus, common features found in the GaSb and InSb nanowires are expected to be observable also in the corresponding [001]-oriented InAs nanowires with a square cross section. Figures~\ref{fig:f001GaSb_Band_Spinor} and \ref{fig:f001InSb_Band_Spinor} show the band structures of the [001]-oriented GaSb and InSb nanowires with a square cross section of size $40\times 40$ nm$^2$ and the spinor distributions of the lowest conduction and highest valence band states in these nanowires. It is seen that the band structures of the two nanowires exhibit the same characteristics. The conduction bands shows simple parabolic dispersions, while the valence bands show non-parabolic, complex dispersions. Du to the square cross section in the nanowires, the systems in the k.p description are $D_{4h}$ symmetric. Thus, all the bands are doubly spin-degenerate and strictly there is no orbital degeneracy allowed in the band structures of the two nanowires. Nevertheless, the 2nd and 3rd lowest conduction bands are very close in energy, forming nearly orbital degenerate bands with the energies at the $\Gamma$-point at
$E\sim 0.84$ eV for the GaSb nanowire and at $E\sim 0.30$ eV for the InSb nanowire. The nearly orbital degeneracy also appears in the 5th and 6th lowest conduction bands with the energies at the $\Gamma$-point at $E \sim 0.87$ eV for the GaSb nanowire and  $E \sim 0.34$ eV for the InSb nanowire, as well as in the the 7th and 8th lowest conduction bands with the energies at the $\Gamma$-point at $E\sim 0.88$ eV for the GaSb nanowire and $E\sim 0.36eV$ for the InSb nanowire. At this point, it should be noted that the orbital degeneracy between the 2nd and 3rd, between the 5th and 6th, as well as between the 7th and 8th lowest conduction bands would appear exactly in the prediction for the band structures of the nanowires with a square cross section based on one-band effective mass theory.  Due to  the presence of the coupling between conduction and valence bands in the narrow band gap semiconductors, this orbital degeneracy is lifted and the orbital degenerate bands should split, although the splittings could be small as we have shown in the results presented above.

In contrast, the valence bands of the [001]-oriented GaSb and InSb nanowires with the square cross sections do not show simple parabolic dispersions, but instead complex characteristics as seen in \reffig{fig:f001GaSb_Band_Spinor} and \reffig{fig:f001InSb_Band_Spinor}.  First, we emphasize again that all the valence bands are doubly spin-degenerate. No nearly orbital degenerate bands are found in the valence bands shown in \reffig{fig:f001GaSb_Band_Spinor} and \reffig{fig:f001InSb_Band_Spinor}. In addition, these valence bands move up or down with increasing $k_z$ vector in a complex way and undergo anti-crossings when they move close to each other in energy. Nevertheless, it should be noted that the all the valence bands are close in energy; the energy separations between them are on the order of $\sim$meV in average, very small when compared to that between the conduction bands which are on the order of $\sim$10 meV. The most distinct feature seen in the valence bands of the [001]-oriented GaSb and InSb nanowires with the square cross sections is a double maximum structure in the topmost valence band of each nanowire with the two energy maxima appearing at finite values of $k_z$. Such a double maximum structure has been seen in the calculations for other [001]-oriented semiconductor nanowires with a square cross section.\cite{2012Peeters6v8,lassen2004exact,Peeters2008optical,kishore2010electronic}. It has also been shown that when the cross section of the [001]-oriented nanowires becomes rectangular, this double maximum structure in the topmost valence bands is suppressed and the bands tend to show good parabolic dispersions  around the $\Gamma$-point.\cite{persson2006electronic,2015Liao}

The band properties of the [001]-oriented GaSb and InSb nanowires can be further analyzed by examining the
the spinor distributions in the band states.  Figures~\ref{fig:f001GaSb_Band_Spinor}(b) and \ref{fig:f001GaSb_Band_Spinor}(d) show the spinor distributions in the lowest conduction band and the highest valence band of the [001]-oriented GaSb nanowire with the square cross section of size $40\times 40$ nm$^2$, while Figs.~\ref{fig:f001InSb_Band_Spinor}(b) and \ref{fig:f001InSb_Band_Spinor}(d) show  the spinor distributions in the lowest conduction band and the highest valence band of the [001]-oriented InSb nanowire with the square cross section of size $40\times 40$ nm$^2$. It is seen that the lowest conduction bands of the two nanowires are dominantly electron-like but do contain some characteristics of hole states.  The extent of the electron-hole state mixing is significantly larger in the InSb nanowire because of a smaller band gap in the material. The highest valence bands of the two nanowires are dominantly characterized by strong mixing of LH and HH states, as it is seen in Figs.~\ref{fig:f001GaSb_Band_Spinor}(d) and \ref{fig:f001InSb_Band_Spinor}(d). Nevertheless, it is found that near the $\Gamma$-point, the highest valence bands appear to be dominantly LH-like. The result can be understood from the localization characterization of the band states as we will discuss below. It is also found in Figs.~\ref{fig:f001GaSb_Band_Spinor}(d) and \ref{fig:f001InSb_Band_Spinor}(d) that with increasing $k_z$, the LH characteristics of the highest valence bands are gradually decreased and the bands switch to be dominantly HH-like at certain values of $k_z$ (near $k_z=0.015$ $\pi /a_0$ in both cases) where the highest valence bands anticross with low-energy-lying HH-like bands.

\begin{figure}[tb]
\includegraphics[width=8.5cm]{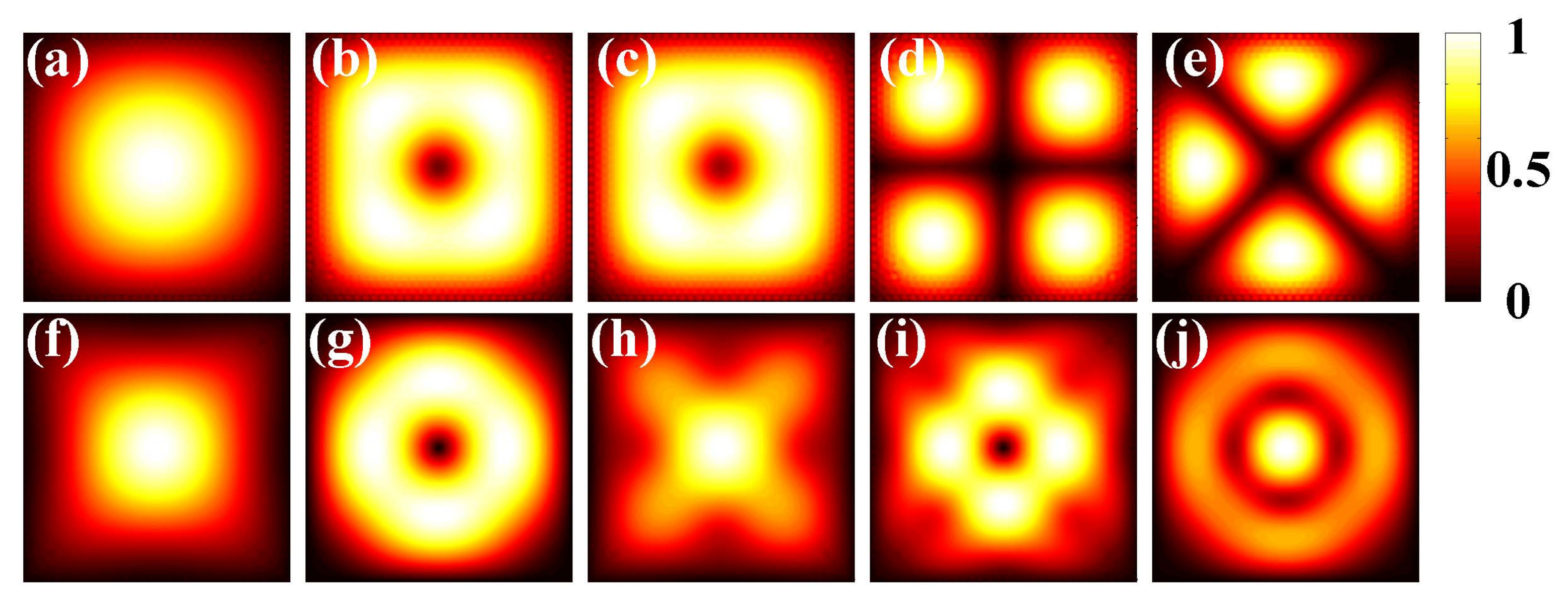}
\caption{Probability distributions of band state wave functions of the [001]-oriented GaSb nanowire with the square cross section of size $l=40$ nm. Panels (a) to (e) show the results for the five lowest conduction band states at the $\Gamma$ point.  Panels (f) to (j) show the results for the five highest valence band states at the $\Gamma$ point.}
\label{fig:f001GaSb_Wave}
\end{figure}

\begin{figure}[tb]
\includegraphics[width=8.5cm]{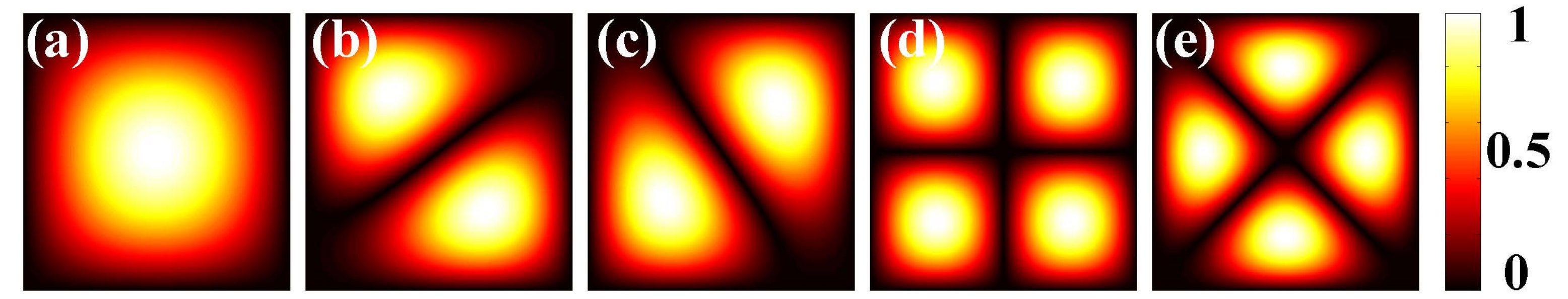}
\caption{The same as in Fig.~\ref{fig:f001GaSb_Wave} but for the probability distributions of the five lowest conduction band states at $\Gamma$ point of the [001]-oriented GaSb nanowire obtained based on simple one-band theory.}
\label{fig:f001GaSb_Wave_Free}
\end{figure}

Now we will discuss the wave functions of the band states of the [001] GaSb and InSb nanowires with the square cross section of size $40\times 40$ nm$^2$. Figure~\ref{fig:f001GaSb_Wave} shows the calculated wave functions for the five lowest conduction band states and the five highest valence band states of the GaSb nanowire at the $\Gamma$ point. Here and throughout this work, the wave functions are presented by the probability distributions normalized within each panel to the highest value found in the panel. It is seen that the lowest conduction band state is $s$-like, while the 2nd and 3rd conduction band states show similar, doughnut-shaped probability distributions. In fact, as can be seen in \reffig{fig:f001GaSb_Band_Spinor}(a), the the 2nd and 3rd conduction bands are very close in energy and form a nearly degenerate band. The 4th lowest conduction band state at the $\Gamma$ point shows four peaks localized at the four corners of the cross section in the probability distribution. The 5th lowest conduction band state also shows a four-peak like probability distribution. But, it is localized at the four edges instead of four corners of the cross section and thus has a higher energy than the 4th lowest conduction band state. In fact, as we discussed in \reffig{fig:f001GaSb_Band_Spinor}(a), this conduction band state is very close in energy to the 6th lowest conduction  band state, forming another nearly degenerate band state. For comparison, we show in \reffig{fig:f001GaSb_Wave_Free} the wave functions of the five lowest conduction band states of the GaSb nanowire at the $\Gamma$-point obtained based on simple one band effective mass theory. It is seen that although the simple one band effective mass theory gives correct descriptions for the wave functions of many conduction band states, it certainly fails to describe the 2nd and   3rd lowest conduction band states.

As for the wave functions of the five highest valence band states at the $\Gamma$-point, it is seen in \reffig{fig:f001GaSb_Wave}(f) to \reffig{fig:f001GaSb_Wave}(j) that they are in general more localized to the inside of the nanowire when compared to their corresponding lowest conduction band states. In particular, the highest valence band, which is also $s$-like, is much more strongly localized to the center of the nanowire than the lowest conduction band state. The 2nd highest valence band state shows a doughnut shaped probability distribution, just as the 2nd lowest conduction band state. This state is more extended to the boundaries of the the nanowire than the highest valence band state and thus feels a stronger quantum confinement. This explains why this dominantly HH-like state has a lower energy than the highest valence band state which is dominantly LH-like. The 3rd, 4th and 5th lowest valence band states, though shaped in more complex patterns in the probability distribution, all are more localized to the inside of the nanowire than their corresponding conduction band states.

\begin{figure}[tb]
\includegraphics[width=8.5cm]{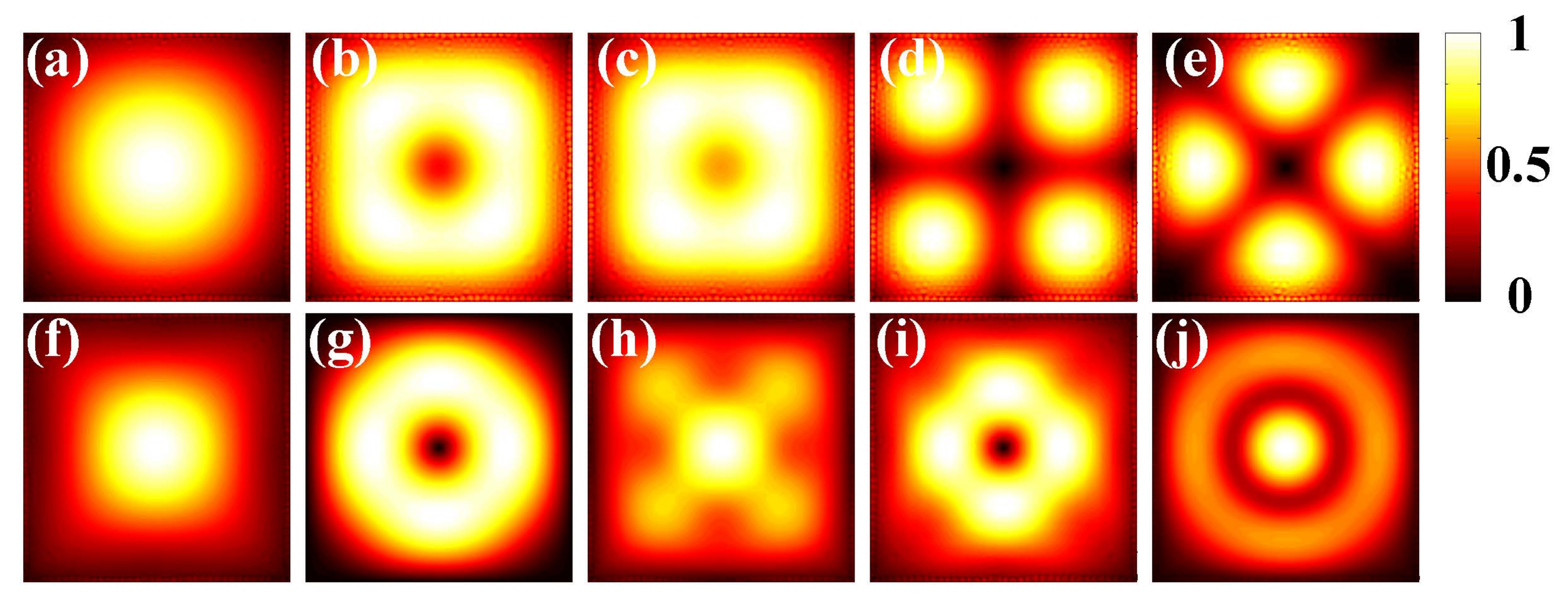}
\caption{Probability distributions of band state wave functions of the [001]-oriented InSb nanowire with the square cross section of size $l=40$ nm. Panels (a) to (e) show the results for the five lowest conduction band states at the $\Gamma$ point.  Panels (f) to (j) show the results for the five highest valence band states at the $\Gamma$ point.}
\label{fig:f001InSb_Wave}
\end{figure}

The wave functions of the five lowest conduction band states and the five highest valence band states at the $\Gamma$ point of the [001]-oriented InSb nanowire with the square cross section of size $40\times 40$ nm$^2$  are shown in Fig.~\ref{fig:f001InSb_Wave}. It is seen that these band states resemble the same probability distribution characteristics as the band states of the GaSb nanowire.

\subsection{InAs nanowires with a square and a circular cross section}

In this subsection, we discuss the band properties of [001]-oriented InAs nanowires. We will consider both the nanowires with a square cross section and the nanowires with a circular cross section. We will first compare the results of calculations for the InAs nanowires with a square cross section to the corresponding GaSb and InSb nanowires shown in Section III A and then discuss the differences in the band properties of the InAs nanowires with a square and with a circular cross section. In the calculations, we have chosen the circular cross section with a radius of 22.57 nm and the square cross section with a size of 40 nm. Thus, the two nanowires have the same cross-sectional areas and therefore similar quantum confinement effects.

\begin{figure}[tb]
\includegraphics[width=8.5cm]{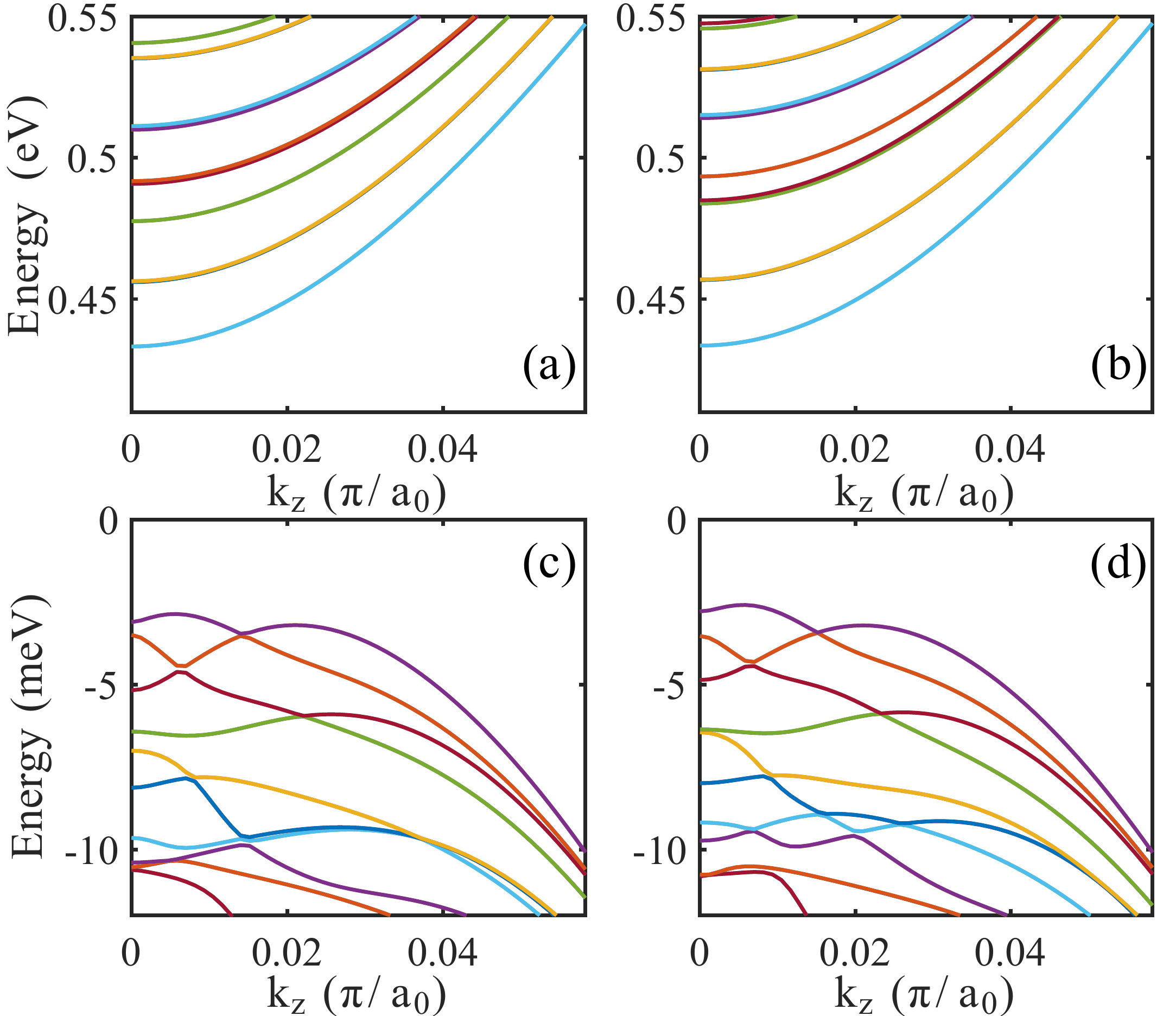}
\caption{(a) and (c) Band structure of an InAs nanowire oriented along the [001] crystallographic direction with a square cross section of size $l=40$ nm. (b) and (d) Band structure of an InAs nanowire oriented along the [001] crystallographic direction with a circular cross section of radius $r=22.57$ nm.}
\label{fig:f001InAs_Band_square_circle}
\end{figure}

Figure~\ref{fig:f001InAs_Band_square_circle} shows the band structures of the [001]-oriented InAs nanowires with a square cross section of size $l=$40 nm [(a) and (c)] and a circular cross-section of radius $r=$22.57 nm [(b) and (d)]. By comparison of the band structure shown in Figs.~\ref{fig:f001InAs_Band_square_circle}(a) and \ref{fig:f001InAs_Band_square_circle}(c) with the band structures shown in Figs.~\ref{fig:f001GaSb_Band_Spinor} and \ref{fig:f001InSb_Band_Spinor}, we see that the [001]-oriented InAs nanowire with the square cross section exhibits the same characteristics in the band structure as the corresponding [001]-oriented GaSb and InSb nanowires, such as formation of nearly degenerate bands in lower conduction bands, a double maximum structure in the topmost valence band, and a complex anti-crossing structure in the valence bands. The differences between the band structures of the three nanowires appear quantitatively in band energies.

Now, we discuss the similarities and differences in the band structures of the [001]-oriented InAs nanowires with the square cross section and the circular cross section. It is seen in \reffig{fig:f001InAs_Band_square_circle} that the band structures of the two nanowires look similar. In particular, the three lowest conduction bands look the same in band shapes and energies (including the nearly degenerate energies of the 2nd and 3rd lowest conduction bands) in the two nanowires. The three highest valence bands of the two nanowires also look much alike. Both nanowires show a double maximum structure in the topmost valence band and complex anti-crossing characteristics between the bands. Nevertheless, clear differences are found in the bands other than the three lowest conduction and the three highest valence bands of the two nanowires. For example, the 4th lowest conduction band in \reffig{fig:f001InAs_Band_square_circle}(a) locates just below 0.48 eV at the $\Gamma$ point and is non-orbital-degenerate band, while the 4th lowest conduction band in \reffig{fig:f001InAs_Band_square_circle}(b) locates above 0.48 eV at the $\Gamma$ point and is nearly orbital-degenerate with the 5th lowest conduction band. For higher conduction bands, this kind of discrepancies persistently appear. The same conclusion can also be made for the valence bands deeper than the 3rd highest valence bands of the two nanowires.

\begin{figure}[tb]
\includegraphics[width=8.5cm]{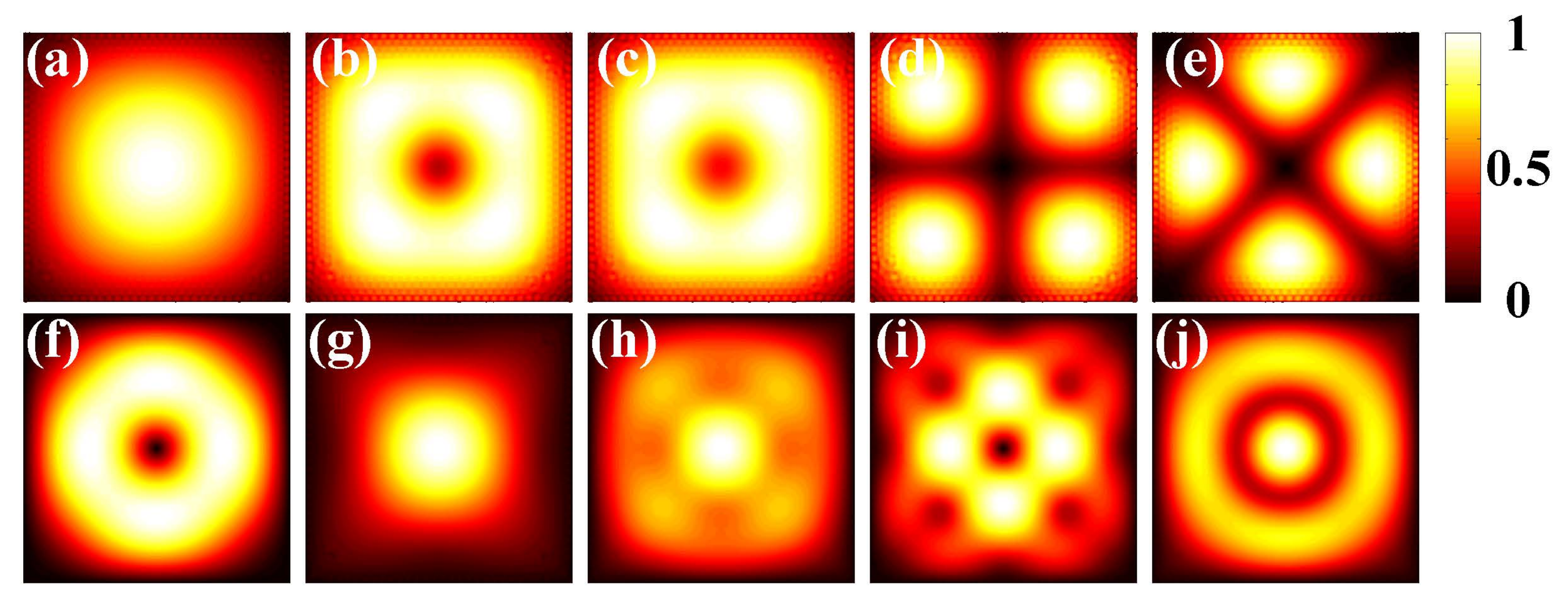}
\caption{Probability distributions of band state wave functions of the [001]-oriented InAs nanowire with the square cross section of size $l=40$ nm. Panels (a) to (e) show the results for the five lowest conduction band states at the $\Gamma$ point.  Panels (f) to (j) show the results for the five highest valence band states at the $\Gamma$ point.}
\label{fig:f001InAs_Wave_Square}
\end{figure}

\begin{figure}[tb]
\includegraphics[width=8.5cm]{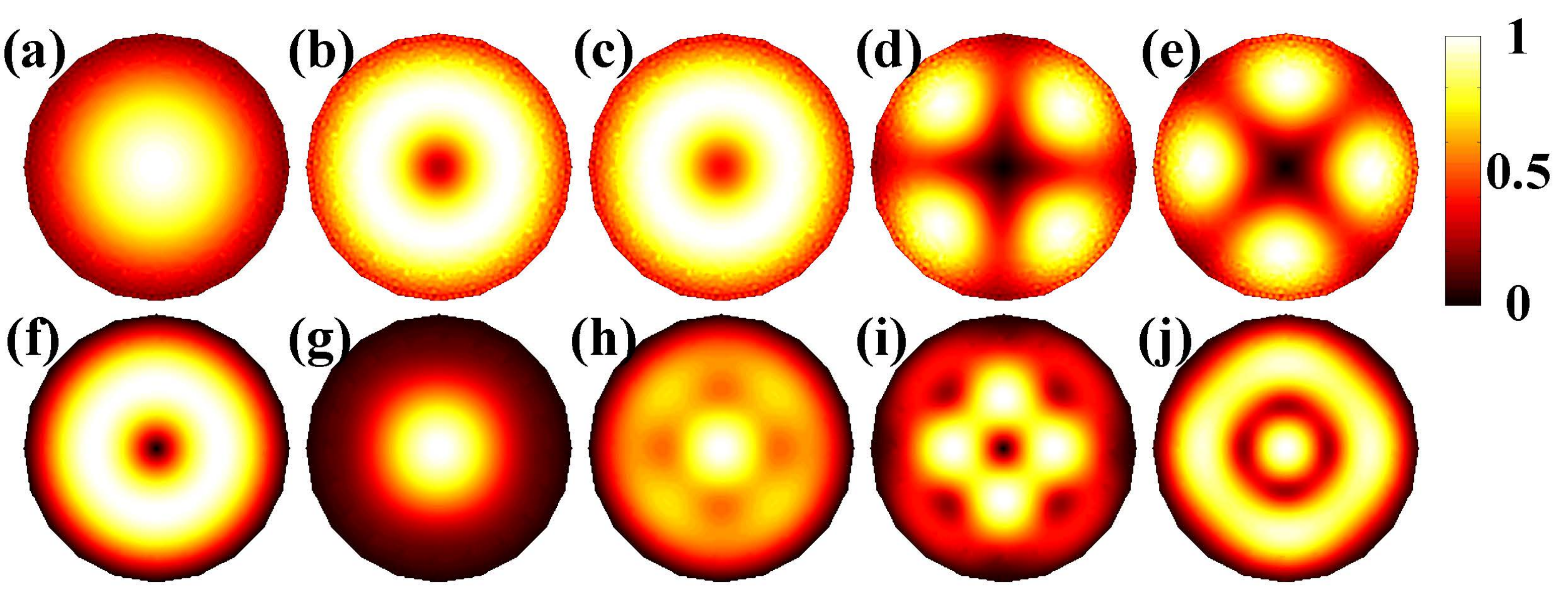}
\caption{Probability distributions of band state wave functions of the [001]-oriented InAs nanowire with the circular section of radius $r=22.57$ nm. Panels (a) to (e) show the results for the five lowest conduction band states at the $\Gamma$ point.  Panels (f) to (j) show the results for the five highest valence band states at the $\Gamma$ point.}
\label{fig:f001InAs_Wave_Circle}
\end{figure}

Figures \ref{fig:f001InAs_Wave_Square} and \ref{fig:f001InAs_Wave_Circle} show the calculated wave functions for the five lowest conduction band  states and the five highest valence band states at the $\Gamma$ point of the [001]-oriented InAs nanowires with the square and the circular cross section. First, for the InAs nanowire with the square cross section, the wave functions of the conduction band states and the valence band states at the $\Gamma$-point look totally similar to those in the corresponding GaSb and InSb nanowires as shown in \reffig{fig:f001GaSb_Wave} and \reffig{fig:f001InSb_Wave}, except for the exchange in the order of the two highest valence states. For the InAs nanowires with the square cross section and the circular cross section, the wave functions of the corresponding states show the same topologies. However, differences are also clearly observable in the wave functions of the band states located far from the band gaps. For example, the 4th and 5th lowest conduction band states of the nanowire with the square cross section show quite different localization properties and thus have very different band energies. As a comparison, the 4th and 5th lowest conduction band states of the nanowire with the circular cross section show almost the same localization properties and thus have nearly the same band energies as we discussed above.
Nevertheless, we would like to emphasize again that it is appropriate to choose either a square cross section or a circular cross section in the analysis of the band properties of a [001]-oriented nanowire if one is interested only in the band properties near the band gap of the nanowire. For the bands located far from the band gap, the effect of the cross section shape on the band properties of the nanowire is not negligible and should be taken into account in the analysis of the electronic properties of the nanowire.

\section{Band structures and band states of [111]-oriented G\lowercase{a}S\lowercase{b}, I\lowercase{n}A\lowercase{s} and I\lowercase{n}S\lowercase{b} nanowires\label{part:111}}

The properties of [111] oriented nanowires are not as widely discussed as the [001] ones, although these nanowires are commonly grown and studied in experiments. The epitaxially grown [111]-oriented III-V semiconductor nanowires often have a hexagonal cross section and the systems in the k.p presentation are $D_{3d}$ symmetric. In addition, the k.p Hamiltonian is symmetric under a combined operation of time reversal and spatial inversion. The energy bands therefore are all doubly degenerate (i.e., spin degenerate), which is the same as we see in the case for the [001]-oriented III-V nanowires. In the following, we will first present and discuss the results obtained for a [111]-oriented GaSb nanowire with a hexagonal cross section and compare them with the results shown above for the [001]-oriented GaSb nanowire. We will then present and discuss, as well as compare, the results of calculations for [111]-oriented InSb nanowires with a hexagonal cross section and a circular cross section. Finally, we present and discuss the results of calculations for [111]-oriented InAs nanowires with a hexagonal cross section of different sizes.

\subsection{GaSb nanowire with a hexagonal cross section}

\begin{figure}[tb]
\includegraphics[width=8.5cm]{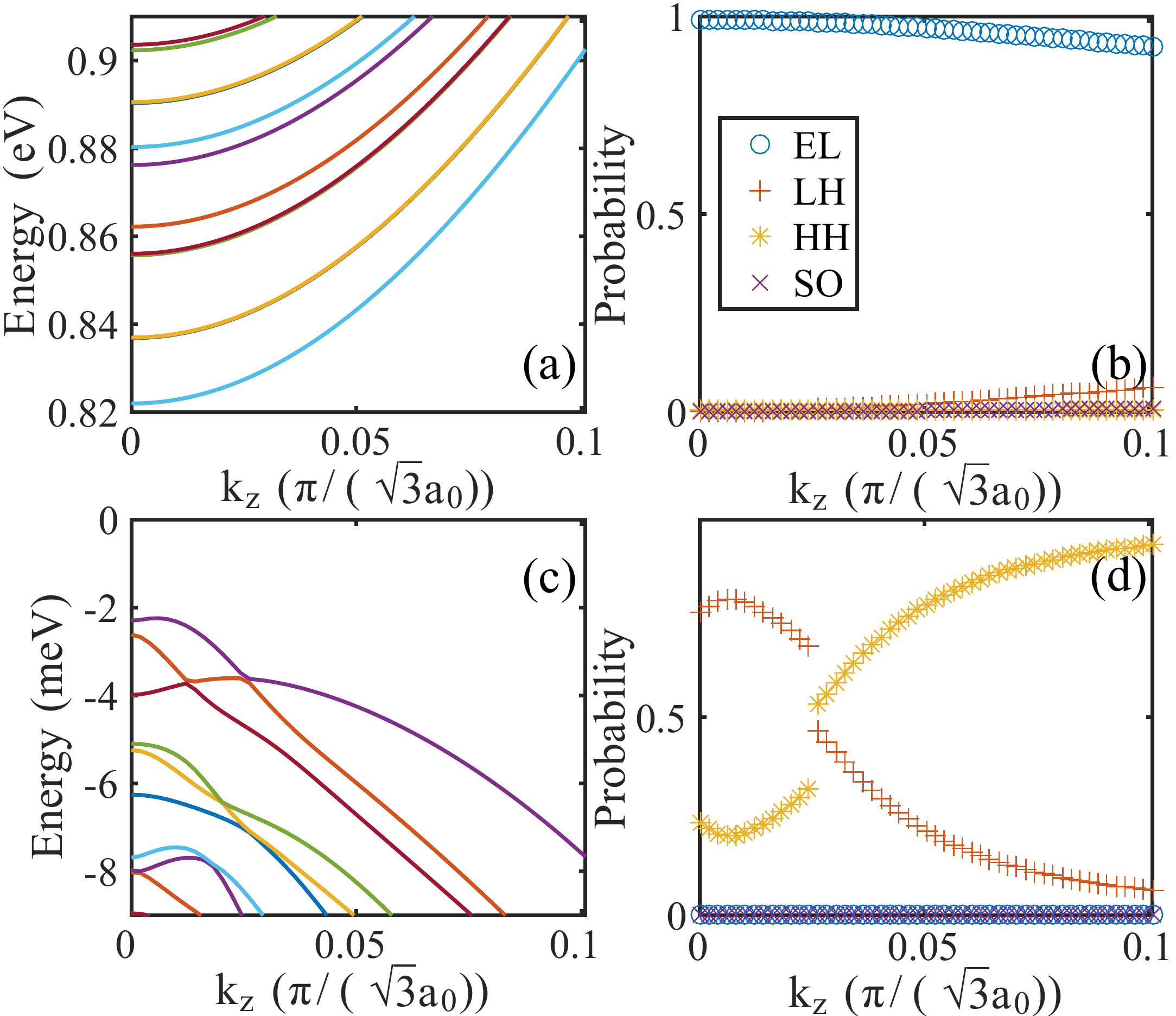}
\caption{(a) and (c) Band structure of a GaSb nanowire oriented along the [111] crystallographic direction with a hexagonal cross section of size $h=24.8$ nm. (b) and (d) Spinor distributions of the lowest conduction band state and the highest valence band state at the $\Gamma$ point of the GaSb nanowire.}
\label{fig:f111GaSb_Band_Spinor}
\end{figure}

Figure~\ref{fig:f111GaSb_Band_Spinor} shows the calculated band structure and spinor distribution of the lowest conduction band and highest valence band of a [111]-oriented GaSb nanowire with a hexagonal cross section of size $h=24.8$ nm (corresponding to a cross-section area which is the same as that in the [001]-oriented GaSb nanowire as we discussed in \reffig{fig:f001GaSb_Band_Spinor}). It is seen that the conduction bands of the [111]-oriented GaSb nanowire show parabolic dispersions just as in the corresponding [001]-oriented nanowire. The lowest conduction band of the [111]-oriented GaSb nanowire is located at $\sim 0.82$ eV and is followed by the nearly orbital-degenerate 2nd and 3rd lowest conduction bands with energies slightly below $0.84$ eV, similar to the case in the [001]-oriented GaSb nanowire. However, not at all like the ones in the [001]-oriented GaSb nanowire, the 4th and 5th lowest conduction bands located slightly below $0.86$ eV are nearly orbital-degenerated in the [111]-oriented GaSb nanowire, while the 6th lowest conduction band laying above $0.86$ eV in the [111]-oriented GaSb nanowire is not. Such differences in nearly orbital degeneracy between the [111]- and [001]-oriented GaSb nanowires appear also in higher energy-laying conduction bands.

As for the valence bands, similar complex, anti-crossing characteristics as seen in the [001]-oriented GaSb nanowire with the square cross section are found in the [111]-oriented GaSb nanowire. However, some noticeable differences are present in the valence bands of the two GaSb nanowires. For example, the highest valence band in the [001]-oriented nanowire does not show the pronounced double-maximum structure as in the [001]-oriented nanowire. Furthermore, although the 2nd highest valence band does show an anti-crossing with the 3rd one, just like the case in the [001]-oriented nanowire as seen in \reffig{fig:f001GaSb_Band_Spinor}(c), it no longer shows an anti-crossing with the 4th highest valence band.

The spinor distributions of the lowest conduction band of the [111]-oriented GaSb nanowire shown in Fig.~\ref{fig:f111GaSb_Band_Spinor}(b) looks the same as for the [001]-oriented GaSb nanowire shown shown in Fig.~\ref{fig:f001GaSb_Band_Spinor}(b). The band is dominantly EL-like and has contained little hole characteristics. The spinor distributions of the highest valence band of the [111]-oriented GaSb nanowire also show great similarities to that in the [001]-oriented GaSb nanowire; the band is dominantly hole-like. In details, the band states are dominantly LH-like states near the $\Gamma$-point and switch to be dominantly HH-like states when passing through the $k_z$ point at which the band shows an anti-crossing with the 3rd highest valence band. Again, the highest valence band contains little characteristics of EL-like and spin SO-like bstates.

\begin{figure}[tb]
\includegraphics[width=8.5cm]{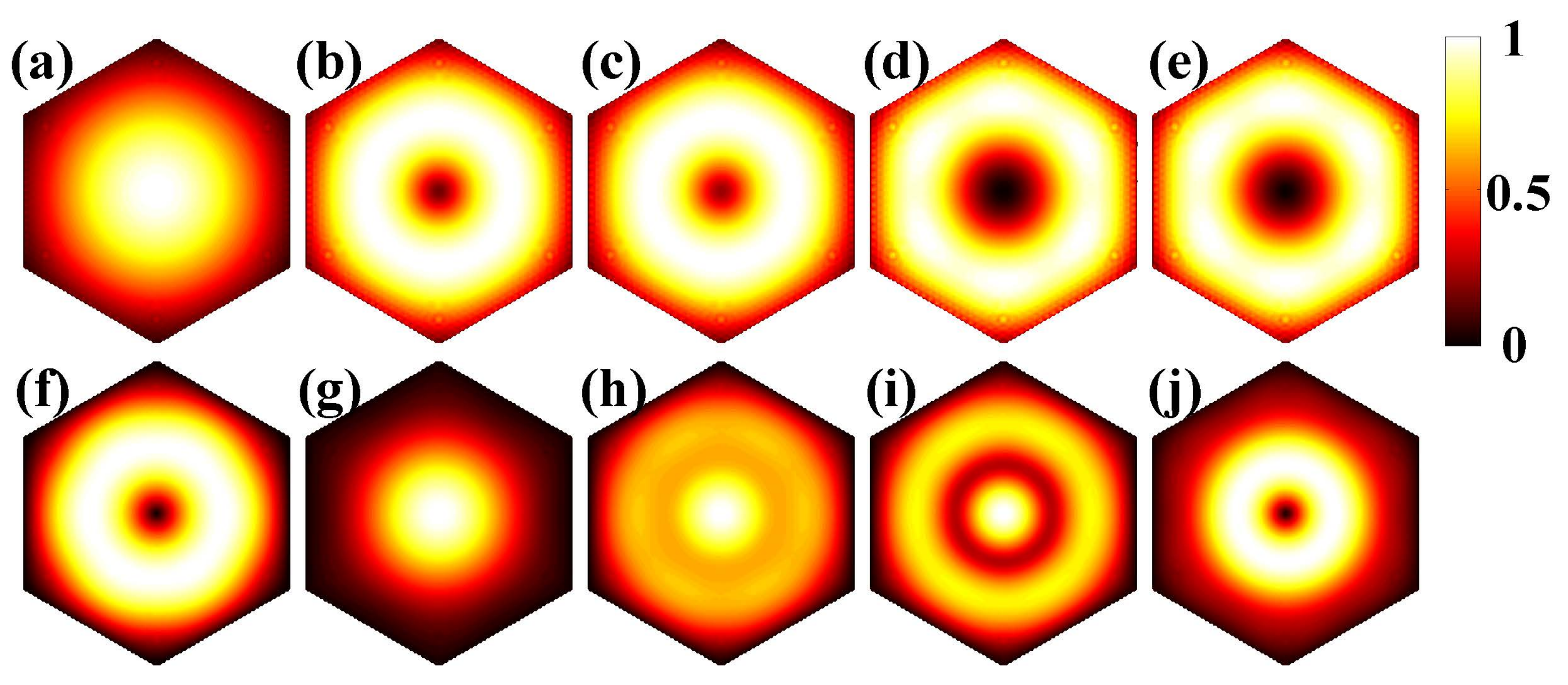}
\caption{Probability distributions of band state wave functions of the [111]-oriented GaSb nanowire with the hexagonal cross section of size $h=24.8$ nm. Panels (a) to (e) show the results for the five lowest conduction band states at the $\Gamma$ point.  Panels (f) to (j) show the results for the five highest valence band states at the $\Gamma$ point.}
\label{fig:f111GaSb_Wave}
\end{figure}

Figure \ref{fig:f111GaSb_Wave} shows the calculated wave functions of the five lowest conduction bands and the five highest valence bands of the [111]-oriented GaSb nanowire at the $\Gamma$ point. When comparing with the wave functions of the band states of the [001]-oriented GaSb nanowire with a square cross section shown in Fig.~\ref{fig:f001GaSb_Wave}, the three lowest conduction band states exhibit similar probability distributions as the three corresponding conduction band states of the [001]-oriented GaSb nanowire. Namely, the lowest conduction band state shows a highly symmetric, circular shaped probability distribution with a peak in the center, while the 2nd and 3rd lowest conduction band states, with nearly degenerated energies, show ringlike probability distributions each with six peaks on the ring and a deep hole in the center. Note that the hole minima of the two conduction band states do not go to zero due to small but not negligible contributions from the valence band states. The next two lowest conduction band state wave functions of the [111]-oriented nanowires show very different probability distributions from the corresponding states in the [001]-oriented GaSb nanowire. These states still exhibit ring-like probability distributions but each with a larger and deeper hole in the middle. 

As for the valence bands of the [111]-oriented GaSb nanowire, the wave functions of the two highest valence band states at the $\Gamma$ point look similar to the two corresponding valence band state wave functions of the [001]-oriented GaSb nanowire, although the ordering of the two states is reversed.  However, the wave function of the next three highest valence band states of the [111]-oriented GaSb nanowire shows distinctly different characteristics in the probability distribution from the three corresponding valence band states of the [001]-oriented GaSb nanowire. The former all show more circularly symmetric probability distributions, the latter exhibit clearly four-fold symmetric probability distributions. Here, it could be worthwhile to note also some details in the three valence band state wave functions of the [111]-oriented GaSb nanowire. It is seen in Fig.~\ref{fig:f111GaSb_Wave} that both the 3rd and 4th highest valence band state wave functions have a circular shaped high probability distribution in the middle and a ring-like high probability distribution around, while the 5th highest valence band state wave function shows a single ring-like probability distribution with a strong localization to the inside of the nanowire.

\subsection{InSb nanowires with a hexagonal and a circular cross section}

\begin{figure}[tb]
\includegraphics[width=8.5cm]{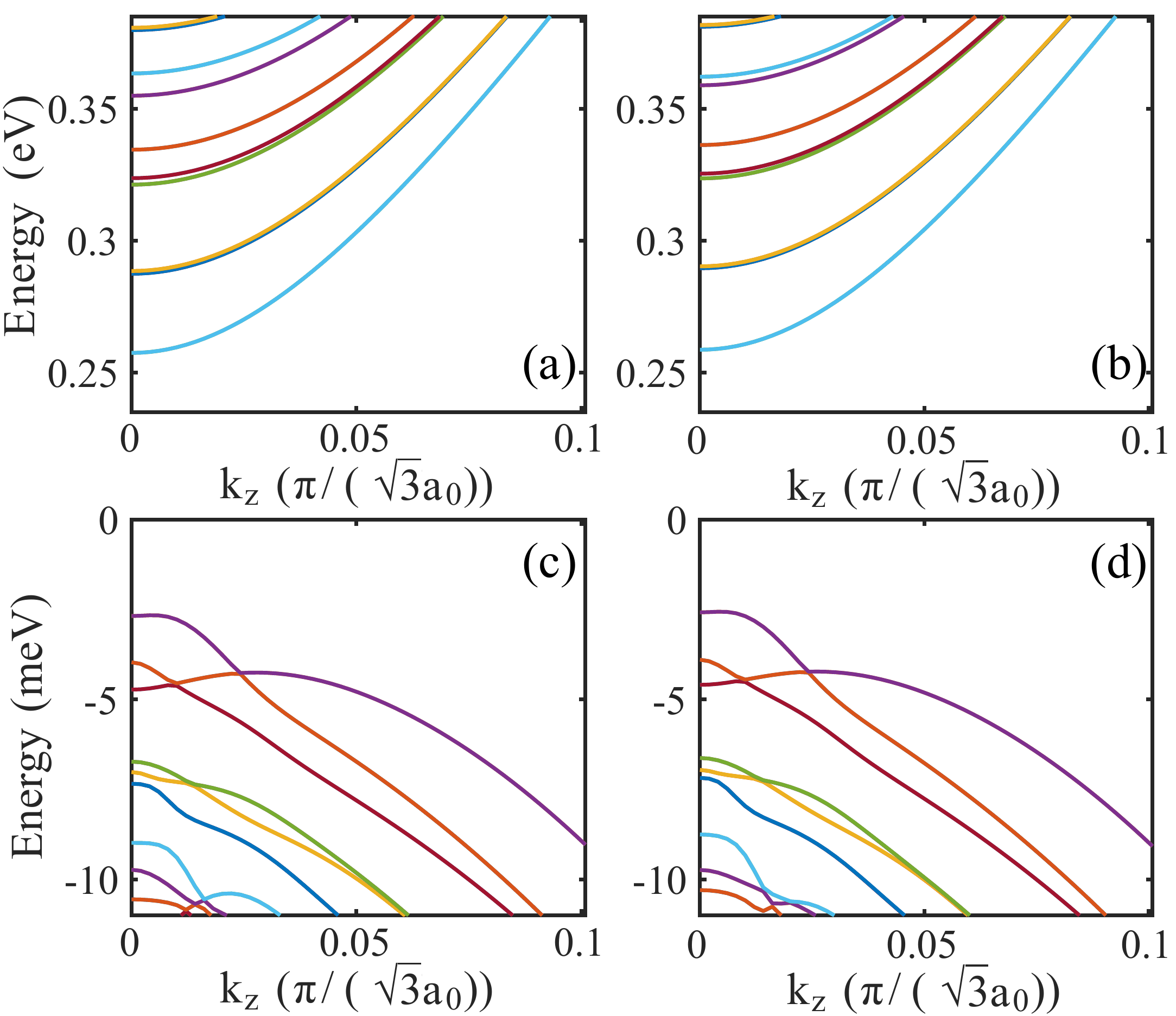}
\caption{(a) and (c) Band structure of an InSb nanowire oriented along the [111] crystallographic direction with a hexagonal cross section of size $h=24.8$ nm. (b) and (d) Band structure of an InSb nanowire oriented along the [111] crystallographic direction with a circular cross section of radius $r=22.57$ nm.}
\label{fig:f111InSb_Band_hex_circle}
\end{figure}

In this section, we present the results of calculations for a [111]-oriented InSb nanowire with a hexagonal cross section of size $h=24.8$ nm and a [111]-oriented InSb nanowire with a circular cross section of radius $r=22.57$ nm (i.e., the two nanowires are the same in cross sectional area) and make a comparable discussion of these results. Figure~\ref{fig:f111InSb_Band_hex_circle} shows the calculated band structures of the two InSb nanowires. In Fig.~\ref{fig:f111InSb_Band_hex_circle}(a), it is seen that the conduction bands of the [111]-oriented InSb nanowire with a hexagon cross-section are all of parabolic type, which is similar to the results found for the [111]-oriented GaSb nanowires shown in Fig.~\ref{fig:f111GaSb_Band_Spinor}, but is different in energy separations between bands due to the fact that the electron effective mass is much smaller in InSb. Again, the lowest conduction band of the [111]-oriented InSb nanowire is non-orbital-degenerate (but spin-degenerate) band, the 2nd and 3rd lowest conduction bands and the 4th and 5th lowest conduction bands form two nearly orbital-degenerate bands, and the 6th band is again a non-orbital-degenerate band. Similarly, as shown in
Fig.~\ref{fig:f111InSb_Band_hex_circle}(b), the valence bands of the [111]-oriented InSb nanowire show complex anti-crossing characteristics as previously seen in the valence bands of the [111]-oriented GaSb nanowire. In  particular,  the three highest valence bands of the [111]-oriented InSb nanowire are all non-orbital-degenerate bands and show the feature that the 3rd highest valence band anti-crosses through the 2nd and 1st highest valence bands as in the [111]-oriented GaSb nanowire. In Figs.~\ref{fig:f111InSb_Band_hex_circle}(b) and \ref{fig:f111InSb_Band_hex_circle}(d), the band structure of the [111]-oriented InSb nanowire with a circular cross section is presented. When compared with that shown in Figs.~\ref{fig:f111InSb_Band_hex_circle}(a) and \ref{fig:f111InSb_Band_hex_circle}(c), little difference is found in the band structures of the two nanowires with the hexagonal and circular cross sections. Thus, all the characteristics of the band structure of the [111]-oriented InSb nanowire with a circular cross section can be described in the same way as for the [111]-oriented InSb nanowire with a corresponding hexagonal cross section.

\begin{figure}[tb]
\includegraphics[width=8.5cm]{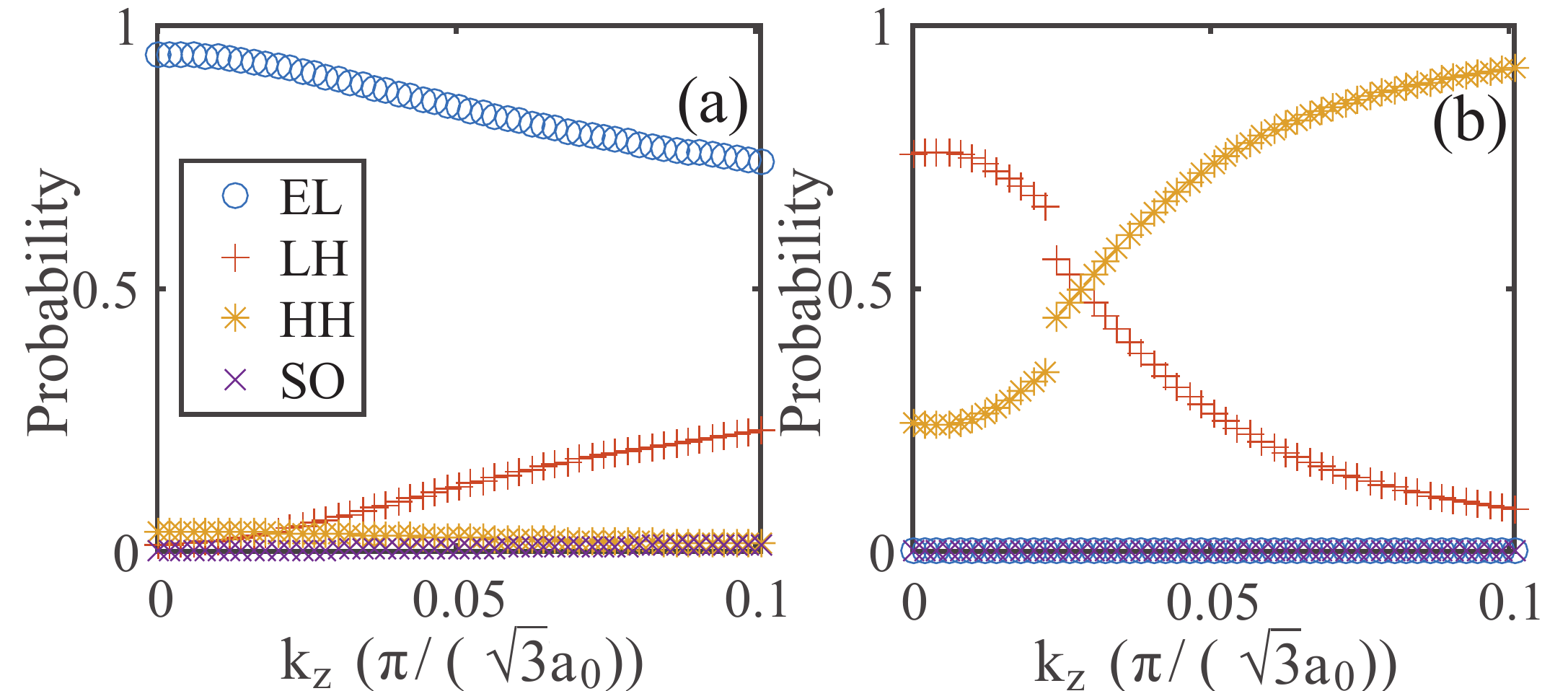}
\caption{Spinor distributions of (a) the lowest conduction band state and (b) the highest valence band state at the $\Gamma$ point of the InSb nanowire oriented along the [111] crystallographic direction with the hexagonal cross section of size $h=24.8$ nm.}
\label{fig:f111InSb_Hex_Spinor}
\end{figure}

The spinor distributions of the lowest conduction band and the highest valence band of the [111]-oriented InSb nanowire with the hexagonal cross section are presented in \reffig{fig:f111InSb_Hex_Spinor}. It is found again  that the conduction band contains significant contributions from the hole-like states as we discussed for the [001]-oriented nanowire shown in \reffig{fig:f001InSb_Band_Spinor}, due to the strong coupling between the electron and hole states in the narrow band gap semiconductor InSb. The HH-like component in the conduction band is larger than the components from the other hole-like states at the $\Gamma$ point, while with increasing $k_z$ the LH-like state contribution increases and becomes the major contribution of the hole-like states. The spinor distributions of the highest valence band again show a strong mixing of the LH- and HH-like states and that the band contains little contribution from the EL-like and spin SO-like states. Also, as the same as in the [111]-oriented GaSb nanowire, the band is dominantly of LH-like characteristics near the $\Gamma$ point and switches to dominantly of HH-like characteristics with increasing $k_z$. The spinor distributions of the lowest conduction band and the highest valence band of the [111]-oriented InSb nanowire with the circular cross section look the same as in \reffig{fig:f111InSb_Hex_Spinor}.

\begin{figure}[tb]
\includegraphics[width=8.5cm]{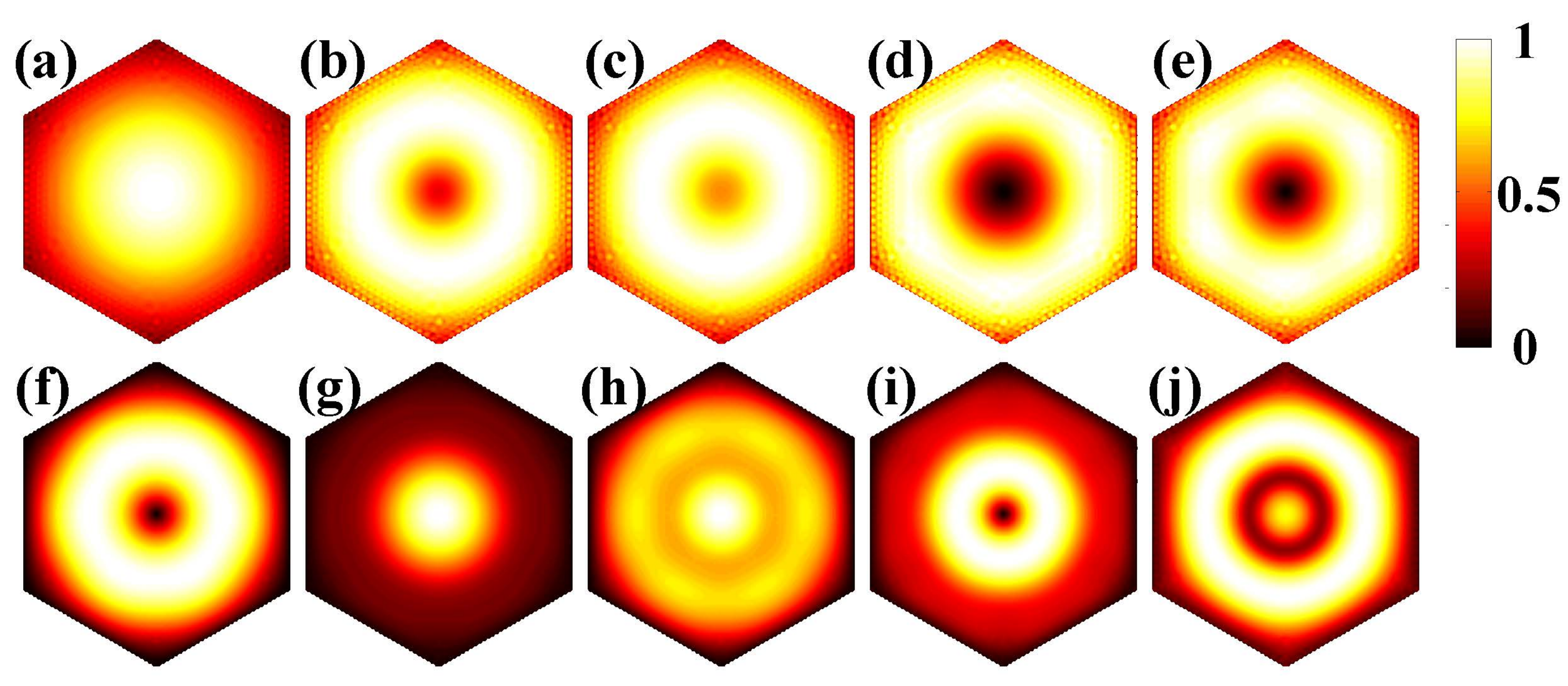}
\caption{Probability distributions of band state wave functions of the [111]-oriented InSb nanowire with the hexagonal cross section of size $h=24.8$ nm. Panels (a) to (e) show the results for the five lowest conduction band states at the $\Gamma$ point.  Panels (f) to (j) show the results for the five highest valence band states at the $\Gamma$ point.}
\label{fig:f111InSb_Wave_Hex}
\end{figure}

Figure \ref{fig:f111InSb_Wave_Hex} shows the wave functions of the five lowest conduction band states and the five highest valence band states at the $\Gamma$ point of the [111]-oriented InSb nanowire with the hexagonal cross section. When compared with the wave functions shown in \reffig{fig:f111GaSb_Wave}, almost the same probability distributions are found in these band states of the [111]-oriented InSb nanowire. One noticeable difference is that the values of the probability distributions of the 2nd and 3rd lowest conduction band states at the center of the InSb nanowire are higher than the corresponding values found in GaSb nanowire, implying a slightly more contributions from the hole states in the nanowire of the narrower band gap InSb material. Another noticeable difference is that the ordering of the probability distributions of the 4th and 5th highest valence band states have been reversed when compared with the corresponding valence band states of the GaSb nanowire. Such an ordering change in the two valence band states have also been found in the calculations based on atomistic tight-binding theory.\cite{2015Liao}

\begin{figure}[tb]
\includegraphics[width=8.5cm]{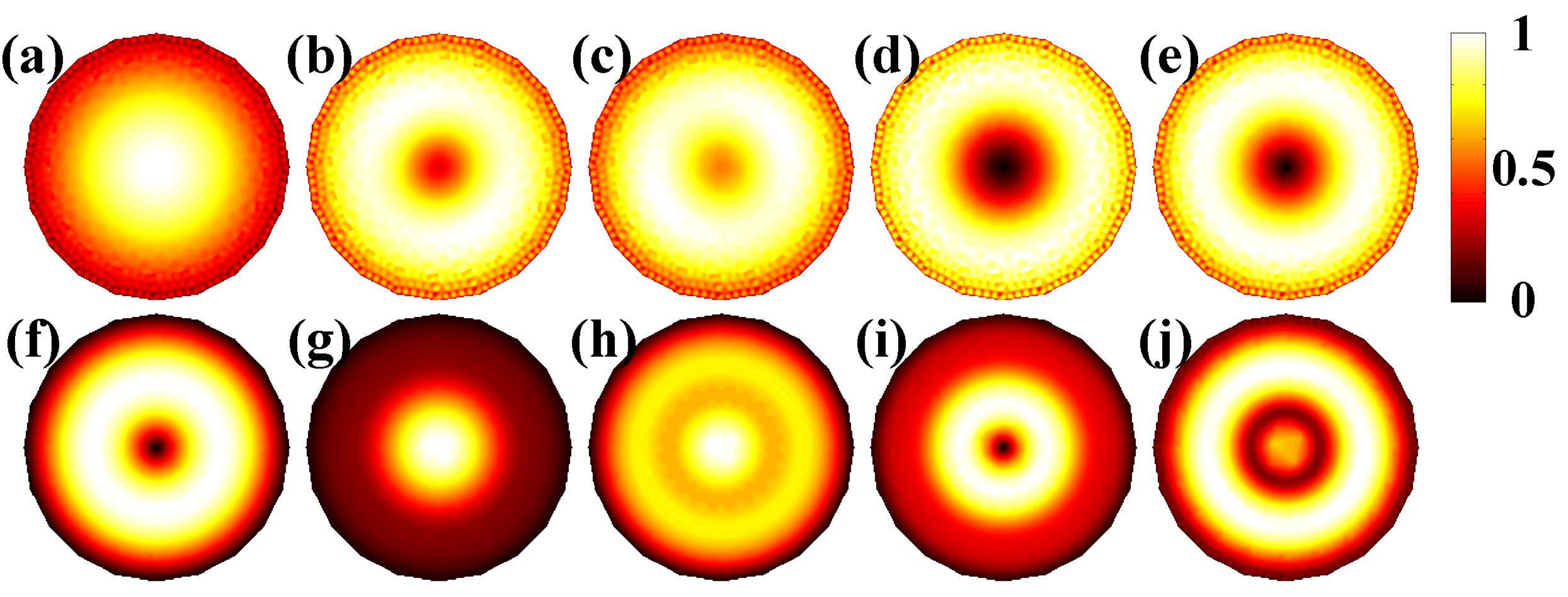}
\caption{Probability distributions of band state wave functions of the [111]-oriented InSb nanowire with the circular cross section of radius $R=22.57$ nm. Panels (a) to (e) show the results for the five lowest conduction band states at the $\Gamma$ point.  Panels (f) to (j) show the results for the five highest valence band states at the $\Gamma$ point.}
\label{fig:f111InSb_Wave_Circle}
\end{figure}

Figure \ref{fig:f111InSb_Wave_Circle} shows the wave functions of the five lowest conduction band states and the five highest valence band states at the $\Gamma$ point of the [111]-oriented InSb nanowire with the circular cross section. It is evident that the probability distributions of the wave functions of these band states look more or less the same as their corresponding states in the InSb nanowire with the hexagonal cross section. The only difference occurs in symmetry details; the $D_{\infty}$ symmetry is found in the probability distributions of the band states of the InSb nanowire with the circular cross section, while the $D_6$ symmetric probability distribution patterns are found for the InSb nanowire with the hexagonal cross section. Nevertheless the difference is very small. Thus, it may well be appropriate to assume a circular cross section in computing for the electronic structure of a [111]-oriented nanowire even if its actual cross section is of hexagonal shape.

\begin{figure*}[tb]
\includegraphics[width=17cm]{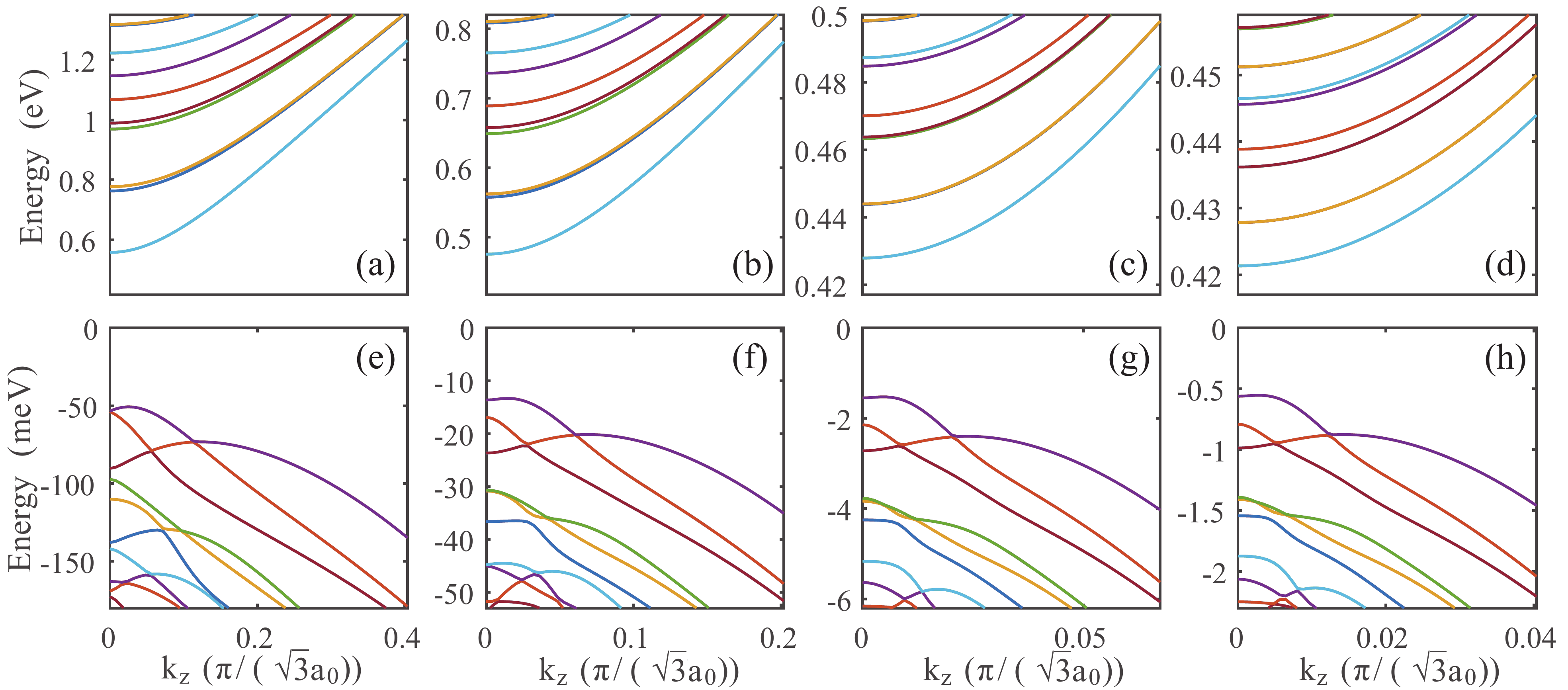}
\caption{Band structure of an InAs nanowire oriented along the [111] crystallographic direction with a hexagonal cross section of (a) and (e) size $h=5$ nm, (b) and (f) size $h=10$ nm, (c) and (g) size $h=30$ nm, and (d) and (h) size $h=50$ nm.}
\label{fig:f111_01_InAs_Band_Series}
\end{figure*}

\begin{figure}[tb]
\includegraphics[width=8.5cm]{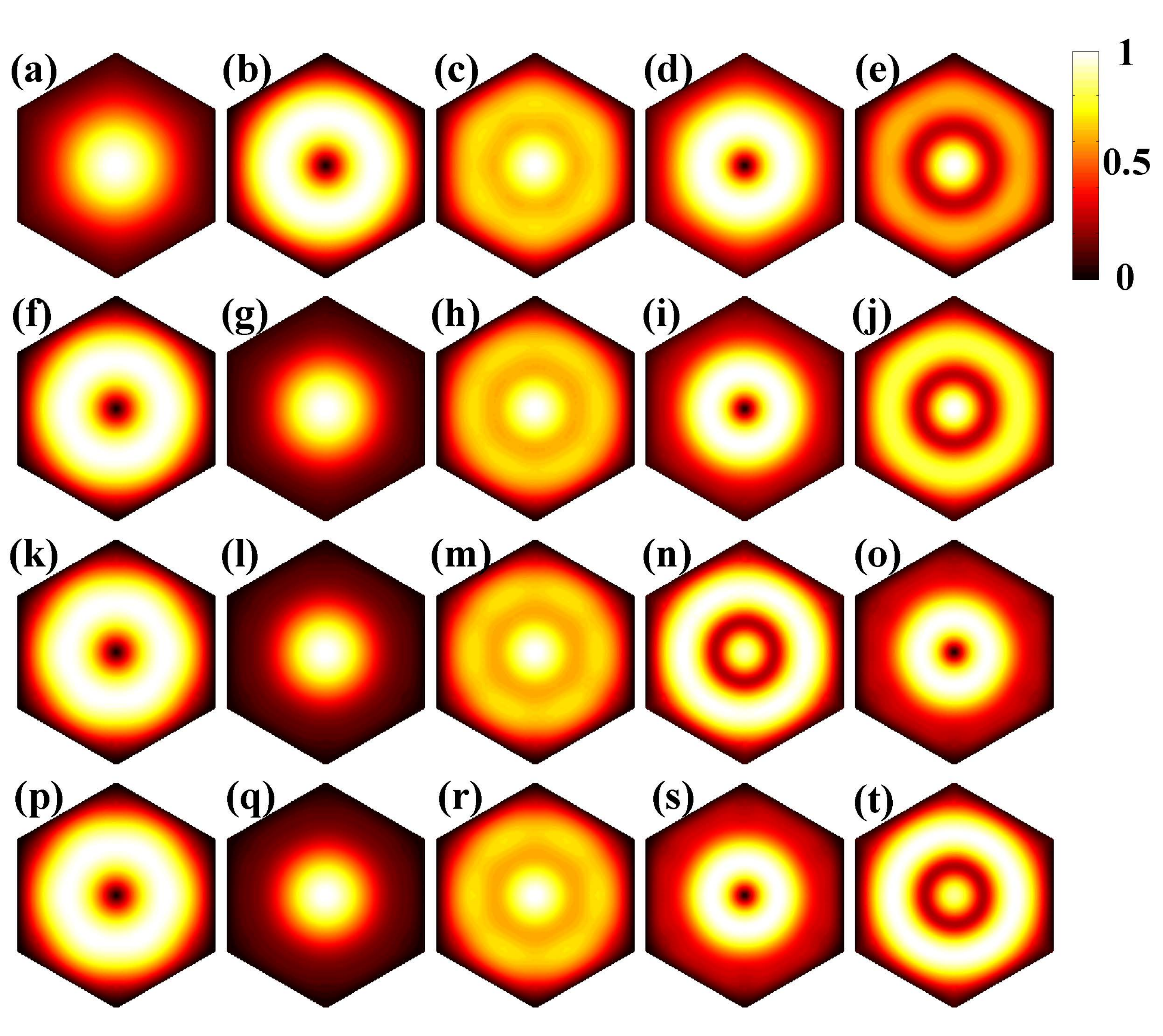}
\caption{Wave function probability distributions of the five highest valence band states of the [111]-oriented InAs nanowire with the hexagonal cross section of (a) to (e) size $h=5$ nm, (f) to (j) size $h=10$ nm, (k) to (o) size $h=30$ nm, and (p) to (t) size $h=50$ nm.}
\label{fig:f111_01_InAs_Wave_Series}
\end{figure}

\subsection{InAs nanowires with hexagonal cross sections of different sizes}

Until now, we have discussed the band properties of the [111]-oriented GaSb and InSb nanowires with a certain cross-sectional size or area. In this subsection,  we will analyze the band properties of nanowires in a series of cross sectional sizes. We take the InAs nanowires with hexagonal cross sections as an example. In \reffig{fig:f111_01_InAs_Band_Series} we plot the band structures of [111]-oriented InAs nanowires with hexagonal cross-sections of 5, 10, 30, and 50 nm in size. It is seen that the four band structures look similar in band shapes but with the band energies scaled with the nanowire cross-sectional size. Such size-scalable characteristics in the band structures of the nanowires were first identified and discussed by Persson and Xu in Ref.~\onlinecite{persson2004electronic}. In detail, the conduction bands all possess good parabolic dispersions and have the same degenerate properties, just like that of the [111]-oriented GaSb and InSb nanowires with a hexagonal cross section. The energy separations between the conduction bands decrease with increasing cross-sectional size in the nanowires. These size-dependent band energies have also been observed and discussed in details in Refs.~\onlinecite{2015Liao} and \onlinecite{2016Liao} and could be evaluated using an empirical formula proposed in Refs.~\onlinecite{2015Liao} and \onlinecite{2016Liao}. As for the valence bands, the general size-scalable band structure characteristics can still be identified in \reffig{fig:f111_01_InAs_Band_Series}. However, we should note that there are visible deviations from the general size-scalable band structure characteristics in the InAs nanowire of small cross-sectional sizes. In particular, the two highest valence bands at the $\Gamma$ point is very close in energy in the InAs nanowire with 5 nm in cross-sectional size, which is clearly different from the InAs nanowires with larger cross-sectional sizes. Differences in valence band shape in the InAs nanowire with 5 nm in cross-sectional size can also be identified when we compare its valence bands to the valence bands of other InAs nanowires with larger cross sectional sizes.

Figure \ref{fig:f111_01_InAs_Wave_Series} shows the wave functions of the five highest valence band states at the $\Gamma$ point of the four [111]-oriented InAs nanowires with different cross-sectional sizes. Here, we do not show the calculated results for the conduction band states, since they look almost the same as for the [111]-oriented GaSb and InSb nanowires shown in Figs.~\ref{fig:f111GaSb_Wave} and \ref{fig:f111InSb_Wave_Hex}. The wave function probability distributions of the corresponding valence band states in the four InAs nanowires of different cross-sectional sizes look very similar. The difference appears in the ordering of these states. For example, the ordering of the wave function probability distributions of the two highest valence band states in the InAs nanowire of 5 nm in cross-sectional size is different from the orderings of the two states in other three InAs nanowires with larger cross-sectional sizes, in agreement with the results shown in \reffig{fig:f111_01_InAs_Band_Series} for the band structures of these nanowires. A switch in the ordering of the wave function probability distributions occurs also in the 4th and 5th highest valence band states of the InAs nanowire of 30 nm in cross-sectional size. This may simply be an effect of numerics, since as can be seen in \reffig{fig:f111_01_InAs_Band_Series} the two valence band states are extremely close in energy in the InAs nanowire with larger cross-sectional sizes.

Finally, we note that we have also calculated the band structures for the [111]-oriented GaSb and InSb nanowires with different cross sectional sizes, and similar size-scalable characteristics in the energy bands and band states have been found.

\section{Conclusions}

In this paper, the electronic structures of narrow band gap semiconductor InSb, InAs and GaSb nanowires oriented along the [001] and [111] crystallographic directions have been studied based on k.p theory. The derivation for a general form of the Luttinger-Kohn eight-band k.p Hamiltonian with the principal axis along an arbitrary crystallographic direction has been presented and a compact form of the Luttinger-Kohn k.p Hamiltonian with the [111] crystallographic direction have been given. The band structures and band states of the nanowires are calculated using FEM. The numerical implementation has been performed in a mixture basis consisting of linear triangular elements and constraint Hermite elements to suppress the Gibbs oscillations  in the eigenstates near the boundaries of the nanowires. Band structures, and band state wave functions and spinor distributions of [001]- and [111]-oriented InSb, InAs, and GaSb are calculated and are compared with respect to the differences in materials, crystallographic orientation and cross-sectional shape. For each orientation, the nanowires of the three narrow band gap materials are found to show qualitatively similar characteristics in the band structures. However, the nanowires oriented along the [001] and [111] directions are found to show different characteristics in the valence bands. In particular, it is found that all the conduction bands show simple, good parabolic dispersions in both the [001]- and [111]-oriented nanowires, while the top valence bands show double-maximum structures in the [001]-oriented nanowires, but single-maximum structures in the [111]-oriented nanowires. The band states near the band gaps are found to be characterized by significant mixtures of electron and hole states in the narrow band gap InSb and InAs nanowires. For the nanowires of GaSb, a material with the largest band gap among the three materials considered, the mixture of electron and hole states in the band states at the band edges is less significant, but is still visible. It is also found that the wave functions of the band states exhibit different probability distribution patterns in the nanowires oriented along the [001] direction and the nanowires oriented along the [111] direction. Although, for the [001]-oriented nanowires with  square cross sections and with circular cross sections, the wave function probability distributions  show some clear differences in certain corresponding band states, such differences become less visible in the [111]-oriented nanowires with a hexagonal cross section and with a circular cross section. It is also shown that single-band effective mass theory could not reproduce all the band state wave functions obtained by the eight-band k.p calculations presented in this work.

\section*{Acknowledgments}

This work was supported by the Ministry of Science and Technology of China (MOST) through the National Basic Research Program 
(Grants No.~2012CB932703 and No.~2012CB932700) and by the National Natural Science Foundation of China (Grants No.~91221202, No.~91421303, and No.~61321001). HQX also acknowledges financial support from the Swedish Research Council (VR).

\section*{Appendix: Numerical implementation based on finite element method}

In this appendix, the numerical implementation in the framework of finite element method (FEM) in solving for the electronic structure of a nanowire described by an eight-band k.p Hamiltonian is presented. In eight-band k.p theory, it is convenient to express the envelope function $F$ as,
\begin{eqnarray}
F=\begin{bmatrix}F_1&F_2&\ldots&F_8\end{bmatrix}^T,
\end{eqnarray}
and the eigenvalue equation in the block form,
\begin{eqnarray}
\sum_{\mu=1}^8H_{\nu\mu}F_{\mu}=EF_{\mu},\quad \nu =1,2,\cdots,8,\label{eq:FEMH}
\end{eqnarray}
where $H_{\nu\mu}$ is an element of the Hamiltonian and $F_{\mu}$ a component of the envelop function. Using the periodic boundary condition for our nanowire system, the above equation can be solved as an eigenvalue problem in the two-dimensional x-y space. Thus, in the following, we will restrict our description to the two-dimensional space. In the framework of FEM, the space spanned by the physical system is divided into a set of mutual exclusive elements $\{e_1,e_2,\cdots\}$, in which a set of real-space basis functions $\{N_{e_1},N_{e_2},\cdots\}$ are defined.
Each envelope function component can be expanded in terms of the basis functions as
\begin{eqnarray}
F_{\mu}=\sum_{i=1}^{M_e} N_{e_i}^TU_{\mu e_i}\label{eq:FEMEXP},
\end{eqnarray}
where $M_e$ is the total number of elements and $\{U_{\mu e_1}~U_{\mu e_2}~\cdots\}$ are the expansion coefficients.
By substituting Eq.~\eqref{eq:FEMEXP} into Eq.~\eqref{eq:FEMH} the eigenvalue equation takes the form
\begin{eqnarray}
\sum_{\mu=1}^8\sum_{i=1}^{M_e} H_{\nu\mu} N^T_{e_i}U_{\mu e_i}=E\sum_{i}^{M_e} N^T_{e_i}U_{\mu e_i}\label{eq:FEM2}.
\end{eqnarray}
By multiplying from left on both sides by $\sum_{j}^{M_e} N_{e_j}$ and integrating over the entire space $\Omega$, the above eigenvalue equation can be written as,
\begin{eqnarray}
\sum_{\mu=1}^8\sum_i^{M_e}\{H_{\nu\mu}\}_{e_i}U_{\mu e_i}=E\sum_i^{M_e}\{M\}_{e_i}U_{\nu e_i}\label{eq:FEMHam},
\end{eqnarray}
where submatrices $\{H_{\nu\mu}\}_{e_i}$ and $\{M\}_{e_i}$ are defined as
\begin{eqnarray}
\{H_{\nu\mu}\}_{e_i}=\iint_{\Omega_{e_i}} N_{e_i}H_{\nu\mu}N^T_{e_i} dxdy,\label{eq:FEMstif1}\\
\{M\}_{e_i}=\iint_{\Omega_{e_i}} N_{e_i}N^T_{e_i} dxdy.\label{eq:FEMstif2}
\end{eqnarray}
Here we note that in the FEM scheme, the basis function $N_{e_i}$ in each element $e_i$ is expressed as a vector obtained by expanding it in a finite and convenient basis, whose dimension and components are given by the type of the element $e_i$ employed (see below for details). Thus, $\{H_{\nu\mu}\}_{e_i}$ and $\{M\}_{e_i}$ in Eqs.~(\ref{eq:FEMstif1}) and (\ref{eq:FEMstif2}) are in general matrices of a finite order.

In actual computation, the above eigenvalue equation [Eq.~(\ref{eq:FEMHam})] is expressed, by assembling the submatrices in all elements  into a larger FEM matrix, as
\begin{align}
[H]&[U]=E[M][U],\label{eq:FEMHamF}\\
[H]&=\begin{bmatrix}
  \{H_{11}\}&\{H_{12}\}&\ldots&\{H_{18}\}\\
  \{H_{21}\}&\{H_{22}\}&\ldots&\{H_{28}\}\\
  \vdots &\vdots &    &\vdots\\
  \{H_{81}\}&\{H_{82}\}&\ldots&\{H_{88}\}\\
\end{bmatrix},\\
[M]&=\begin{bmatrix}
  \{M\}&&&&\\
  &\{M\}&&&\\
  &&\ddots&\\
  &&&&\{M\}\\
\end{bmatrix},\label{eq:M} \\
[U]&=\begin{bmatrix}U_1&U_2&\ldots&U_8\end{bmatrix}^T.
\end{align}
where $U_i$ is an eigenvector component, which itself is a vector of a finite dimension.  This equation needs to be solved under the boundary condition of
\begin{equation}
F_{\mu}(x,y)|_{\text{cross-section boundary}}=0,\text{ for } \mu= 1\cdots 8.
\end{equation}
Practically, this can be achieved by solving the eigenvalue equation with matrices of a reduced size in Eq.~(\ref{eq:FEMHamF}), obtained by eliminating the rows and columns whose corresponding components in the eigenvectors are known to be zero according to the boundary condition.

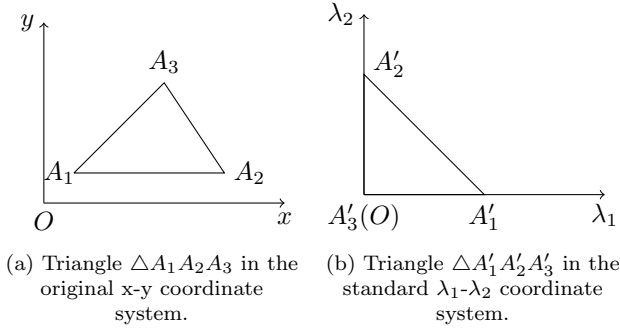
\begin{figure}[tb]
\centering
\subfigure[\label{subfig:tri_or} Triangle $\triangle A_1A_2A_3$ in the original x-y coordinate system.] {
\begin{tikzpicture}[scale=0.80]
  \draw[->] (0,0) node[below] {$O$} -- (4,0) node[below] {$x$};
  \draw[->] (0,0) -- (0,3) node[left]{$y$};
  \draw (0.5,0.5) node[left=-4pt]{$A_1$}--(3,0.5) node[right]{$A_2$}--(2,2) node[above]{$A_3$}--(0.5,0.5);
\end{tikzpicture}}
\subfigure[\label{subfig:tri_sd} Triangle $\triangle A'_1A'_2A'_3$ in the standard $\lambda_1$-$\lambda_2$ coordinate system.] {
\begin{tikzpicture}[scale=0.80]
 \draw[->] (0,0) -- (4,0) node[below] {$\lambda_1$};
 \draw[->] (0,0) -- (0,3) node[left]{$\lambda_2$};
 \draw (0,0) node[below]{$A'_3(O)$}--(2,0) node[below]{$A'_1$}--(0,2) node[above=4pt,right]{$A'_2$}--(0,0);
\end{tikzpicture}
}
\caption{Transformation of (a) a triangle in the origin x-y coordinate system to (b) a triangle in the standard $\lambda_1$-$\lambda_2$ coordinate system.}
\label{Fig:transformation}
\end{figure}

Solving for the above FEM eigenvalue equation can be implemented only if the specific expressions of the basis functions $N_{e_i}$ in all elements $e_i$ are specified. To provide these expressions, it is convenient to transform the original x-y coordinate system to a standard $\lambda_1$-$\lambda_2$ coordinate system defined by
\begin{eqnarray}
x&=&(x_1-x_3)\lambda_1+(x_2-x_3)\lambda_2+x_3\label{eq:FEMxtol1},\\
y&=&(y_1-y_3)\lambda_1+(y_2-y_3)\lambda_2+y_3\label{eq:FEMxtol2},
\end{eqnarray}
as shown in \reffig{Fig:transformation}, where a triangle $\triangle A_1A_2A_3$ in the x-y plane defined with coordinates $(x_1,y_1)$, $(x_2,y_2)$, and $(x_3,y_3)$ has been transformed to a triangle $\triangle A'_1A'_2A'_3$ defined with the corresponding coordinates of $(1,0)$, $(0,1)$, and $(0,0)$ in the $\lambda_1$-$\lambda_2$ plane.

The transformations of the derivatives in the two coordinate systems can be obtained from the following  relations,
\begin{eqnarray}
\frac{\partial}{\partial x}&=&\frac{1}{\det A}\big[(y_2-y_3)\frac{\partial}{\partial \lambda_1}-(y_1-y_3)\frac{\partial}{\partial \lambda_2}\big],\\
\frac{\partial}{\partial y}&=&\frac{1}{\det A}\big[-(x_2-x_3)\frac{\partial}{\partial \lambda_1}+(x_1-x_3)\frac{\partial}{\partial \lambda_2}\big],
\end{eqnarray}
where
\begin{eqnarray}
A&=&\begin{bmatrix}1&1&1\\x_1&x_2&x_3\\y_1&y_2&y_3\end{bmatrix}.
\end{eqnarray}
Thus, operators in the k.p Hamiltonian, such as $C\hat{k}^2_x$, $C\hat{k}^2_y$, $C\hat{k}_x$, $C\hat{k}_y$, $C\hat{k}_x\hat{k}_y$, $C\hat{I}$, where $C$ stands for a material parameter, are transformed according to
\small
\begin{align}
\{C\hat{k}^2_x\}=&-[\frac{\partial}{\partial x}C\frac{\partial}{\partial x}]\nonumber\\
=&\frac{C}{\det A}T^T\bigg[(y_2-y_3)^2\langle B_1B_1^T\rangle -(y_2-y_3)(y_1-y_3)\nonumber\\
&\times (\langle B_2B_1^T\rangle +\langle B_1B_2^T\rangle )+(y_1-y_3)^2\langle B_2B_2^T\rangle \bigg]T,\\
\{C\hat{k}^2_y\}=&-[\frac{\partial}{\partial y}C\frac{\partial}{\partial y}]\nonumber\\
=&\frac{C}{\det A}T^T\bigg[(x_2-x_3)^2\langle B_1B_1^T\rangle -(x_2-x_3)(x_1-x_3)\nonumber\\
&\times (\langle B_2B_1^T\rangle +\langle B_1B_2^T\rangle )+(x_1-x_3)^2\langle B_2B_2^T\rangle \bigg]T,\\
\{C\hat{k}_x\}=&-\frac{i}{2}[C\frac{\partial}{\partial x}+\frac{\partial}{\partial x}C]\nonumber\\
=&-\frac{iC}{2}T^T\bigg[(y_2-y_3)(\langle N_{e_i}B_1^T\rangle -\langle B_1N_{e_i}^T\rangle )-\nonumber\\
&(y_1-y_3)(\langle N_{e_i}B_2^T\rangle -\langle B_2N_{e_i}^T\rangle )\bigg]T,\\
\{C\hat{k}_y\}=&-\frac{i}{2}[C\frac{\partial}{\partial y}+\frac{\partial}{\partial y}C]\nonumber\\
=&-\frac{iC}{2}T^T\bigg[-(x_2-x_3)(\langle N_{e_i}B_1^T\rangle -\langle B_1N_{e_i}^T\rangle )+\nonumber\\
&(x_1-x_3)(\langle N_{e_i}B_2^T\rangle -\langle B_2N_{e_i}^T\rangle )\bigg]T,\\
\{C\hat{k}_x\hat{k}_y\}=&-\frac{1}{2}[\frac{\partial}{\partial x}C\frac{\partial}{\partial y}+\frac{\partial}{\partial y}C\frac{\partial}{\partial x}]\nonumber\\
=&\frac{C}{2\det A}T^T\bigg[-2(x_2-x_3)(y_2-y_3)\langle B_1B_1^T\rangle\nonumber\\
+&\big((x_1-x_3)(y_2-y_3)+(y_1-y_3)(x_2-x_3)\big)\langle B_2B_1^T\rangle\nonumber\\
+&\big((x_2-x_3)(y_1-y_3)+(y_2-y_3)(x_1-x_3)\big)\langle B_1B_2^T\rangle \nonumber\\
&-2(x_1-x_3)(y_1-y_3)\langle B_2B_2^T\rangle \bigg]T,\\
\{C\hat{I}\}=&(C\cdot\det A)\cdot T^T\langle N_{e_i}N^T_{e_i}\rangle T,
\end{align}
\normalsize
where $N_{e_i}$ is the basis function in element $e_i$ defined in the $\lambda_1$-$\lambda_2$ coordinate system, $T$ an element-type related matrix, $B_1=\frac{\partial}{\partial \lambda_1}N_{e_i}$, $B_2=\frac{\partial}{\partial \lambda_2}N_{e_i}$, and angle braces for any given expression inside stand for integration operations as
\begin{equation}
\langle AB^T\rangle =\iint_{\Omega_{e_i}}AB^Tdxdy.\label{Integral}
\end{equation}
In the linear triangular elements, the basis function $N_{e_i}$ and matrix $T$ take the forms of
\begin{align}
N_{e_i}&=\begin{bmatrix}\lambda_1&\lambda_2&\lambda_3\end{bmatrix}^T,\\
T&=diag\{1,1,1\},
\end{align}
where $\lambda_3$ is given by $\lambda_3=1-\lambda_1-\lambda_2$. In the constraint Hermite triangular elements, $N_{e_i}$ and $T$ are given by
\begin{align}
N_e&=\begin{bmatrix}Q_1&Q_2&Q_3&R_1&S_1&R_2&S_2&R_3&S_3\end{bmatrix}^T,\\
T&=diag\{1,1,1,D,D,D\},
\end{align}
where
\begin{align}
D&=\begin{bmatrix}x_1-x_3&y_1-y_3\\x_2-x_3&y_2-y_3\end{bmatrix},\\
\nonumber\\
Q_1(\lambda)&=\lambda_1^2(3-2\lambda_1)+2\lambda_1\lambda_2\lambda_3,\nonumber\\
Q_2(\lambda)&=\lambda_2^2(3-2\lambda_2)+2\lambda_1\lambda_2\lambda_3,\nonumber\\
Q_3(\lambda)&=\lambda_3^2(3-2\lambda_3)+2\lambda_1\lambda_2\lambda_3,\nonumber\\
R_1(\lambda)&=\lambda_1^2(\lambda_1-1)-\lambda_1\lambda_2\lambda_3,\nonumber\\
R_2(\lambda)&=\lambda_1\lambda_2^2+\frac{1}{2}\lambda_1\lambda_2\lambda_3,\nonumber\\
R_3(\lambda)&=\lambda_1\lambda_3^2+\frac{1}{2}\lambda_1\lambda_2\lambda_3,\nonumber\\
S_1(\lambda)&=\lambda_1^2\lambda_2+\frac{1}{2}\lambda_1\lambda_2\lambda_3,\nonumber\\
S_2(\lambda)&=\lambda_2^2(\lambda_2-1)-\lambda_1\lambda_2\lambda_3,\nonumber\\
S_3(\lambda)&=\lambda_2\lambda_3^2+\frac{1}{2}\lambda_1\lambda_2\lambda_3.
\end{align}
Finally, we note that with the above expressions for $N_{e_i}$, the integration of Eq.~(\ref{Integral}) can be easily evaluated using the elegant formula of
\begin{equation}
\iint_{\Omega_{e_i}}\lambda_1^n\lambda_2^p\lambda_3^q dxdy=\frac{n!p!q!}{(n+p+q)!}\det A .
\end{equation}

\end{document}